\documentclass[11pt,a4paper]{article}
\usepackage[pdftex]{graphicx}
\usepackage[numbers,sort&compress]{natbib}
\usepackage{multirow}
\usepackage{amssymb}
\usepackage{amsmath,amsfonts}
\usepackage[symbol]{footmisc}
\usepackage{url}
\usepackage{longtable}
\usepackage[hyperfootnotes=false]{hyperref}

\newcommand{\CELL}{\texttt{CELL}}
\newcommand{\eg}{{\it e.g.}, }
\newcommand{\ie}{{\it i.e.}, }

\renewcommand{\text}[1]{%
{\mbox{\tiny #1}}%
}

\DeclareMathDelimiter{\lVert}
{\mathopen}{symbols}{"6B}{largesymbols}{"0D}
\DeclareMathDelimiter{\rVert}
{\mathclose}{symbols}{"6B}{largesymbols}{"0D}
\usepackage{listings}
\usepackage{xcolor}
\definecolor{mauve}{rgb}{0.2, 0.5, 0.478}
\definecolor{dkgreen}{rgb}{0.2, 0.5, 0.478}
\usepackage{bm}
\usepackage{graphicx}
\usepackage{fancyvrb}
\usepackage{multirow}
\usepackage{url}
\begin{document}

\parskip 1ex



\begin{center}
{\Large \bf \CELL: a Python package for cluster expansion with a focus on complex alloys}

{Santiago Rigamonti$\footnote[1]{srigamonti@physik.hu-berlin.de}$, Maria Troppenz, Martin Kuban, Axel H\"ubner, and Claudia Draxl}

{\it\small Humboldt-Universit\"at zu Berlin, Institut f\"ur Physik and IRIS Adlershof, 12489 Berlin, Germany}\\

\vspace{0.2cm}
{\small 27 October 2023}

\end{center}

\vspace{0.1cm}
\vspace{0.1cm}

\begin{abstract}
We present the Python package \texttt{CELL}, which provides a modular approach to the cluster expansion (CE) method. \texttt{CELL} can treat a wide variety of substitutional systems, including one-, two-, and three-dimensional alloys, in a general multi-component and multi-sublattice framework. It is capable of dealing with complex materials comprising several atoms in their \textit{parent lattice}. \texttt{CELL} uses state-of-the-art techniques for the construction of training data sets, model selection, and finite-temperature simulations. The user interface consists of well-documented Python classes and modules (http://sol.physik.hu-berlin.de/cell/). \texttt{CELL} also provides visualization utilities and can be interfaced with virtually any \textit{ab initio} package, total-energy codes based on interatomic potentials, and more. The usage and capabilities of \texttt{CELL} are illustrated by a number of examples, comprising a Cu-Pt surface alloy with oxygen adsorption, featuring two coupled binary sublattices, and the thermodynamic analysis of its order-disorder transition; the demixing transition and lattice-constant bowing of the Si-Ge alloy; and an iterative CE approach for a complex clathrate compound with a parent lattice consisting of 54 atoms.
\end{abstract}

\vspace{0.5cm}

\section{Introduction}
Typical quests in materials science, as for instance finding stable compositions of an alloy and its properties, or determining the conditions for molecular adsorption on a surface, involve computations of a large number of atomic configurations on a well-defined lattice. Ideally, one would perform these computations using first-principles methods, \eg density-functional theory (DFT), however, due to the combinatorial explosion of the number of configurations with system size, a direct \textit{ab initio} approach is in many cases out of reach. In this context, the cluster expansion (CE) method \cite{Connolly1983,Sanchez19840} can be used to achieve a physical description of materials with essentially \textit{ab initio} accuracy at reasonable computational cost. In this method, the configuration-dependent properties of the material are computed by means of generalized Ising-like models parametrized with \textit{ab initio} data. Thus, calculations of a huge number of atomic configurations with large supercells become feasible, bridging length scales and enabling a statistical-thermodynamics description.

Despite these capabilities, systems of technological interest are often still too complex. Dealing with complexity requires adequate code, able to account for multi-component settings, multiple sublattices, large parent lattices, surfaces and interfaces, and more. Moreover, CE modeling also entails tasks like \textit{learning from data}, including the creation of data sets for model training and the evaluation and selection of models. In this context, the application of modern artificial intelligence (AI) techniques to CE model construction is possible and desirable. In this work, we present the CE Python \cite{Van-Rossum2009} package \texttt{CELL} \cite{Cell2019} that provides a modular approach to cluster expansion, fulfilling all the needs mentioned above. \texttt{CELL} allows the set-up of systems with an arbitrary number of substituent species types and an arbitrary number of sublattices. Here, a sublattice is a subset of crystal sites whose composition can differ from other sites. The size of the parent lattice, \ie the number of atoms comprising the primitive unit cell of the pristine (non-substituted) material, can be arbitrarily large, so that \texttt{CELL} can readily deal with complex alloys having several atoms in the primitive unit cell. One, two, and three dimensions can be handled, which allows, for instance, the study of surface alloys. 

The construction of CE models involves fitting to, \eg \textit{ab initio} data sets. Since these data are often very costly to compute, \texttt{CELL} implements a number of structure selection strategies aimed at rationalizing data sets for optimal model training. These include special quasirandom structures \cite{Zunger1990} and variance-reduction schemes \cite{Walle2002,Mueller2010}. \texttt{CELL} is written in Python \cite{Van-Rossum1995}, and provides a common interface to the various estimators from the machine-learning Python library \texttt{scikit-learn} \cite{Pedregosa2011,Buitinck2013}, and to native estimators, \ie those coded inside the \texttt{CELL} package. Its modularity allows for the \textit{ad hoc} design and implementation of AI strategies. At the core of the representation of crystal structures, the \texttt{Atoms} object of the Atomic Simulation Enviroment (ASE)  \cite{Larsen2017} is employed, giving access to all the advantages of ASE, as for instance the construction of structure databases and the interface to a large number of \textit{ab initio} DFT codes. \texttt{CELL} provides state-of-the-art tools for performing thermodynamic analysis of materials. To this extent, a module for the calculation of the configurational density of states with the Wang-Landau~\cite{Borg2005} method is available. This module can be executed in parallel, which paves the way for performing simulations of very large supercells.

The paper is organized as follows: Sec.~\ref{sec:ce-gen} gives a general introduction to the CE method; Sec.~\ref{sec:ce} and Sec.~\ref{sec:thermo} present the structure of the code for model construction and thermodynamical analysis, respectively, along with an illustration of its most important features through a complex surface system consisting of a Pt/Cu(111) surface alloy with atomic O adsorption. Finally, the application to SiGe, and the complex thermoelectric clathrate compound Ba$_8$Al$_x$Si$_{46-x}$, with $54$ atoms in the primitive cell, are presented in Sec.~\ref{sec:apps}. Conclusions are given in Sec.~\ref{sec:con}.

\section{Cluster expansion method\label{sec:ce-gen}}
In this section, we present an overview of the cluster expansion formalism for the multi-component and multi-sublattice case \cite{Sanchez1984,Walle2009}. Additionally, formal aspects related to the application of AI techniques to CE model construction are introduced. 

\subsection{Simple CE of a binary alloy}
\begin{figure}[htpb]
\centering
\includegraphics[scale=0.30]{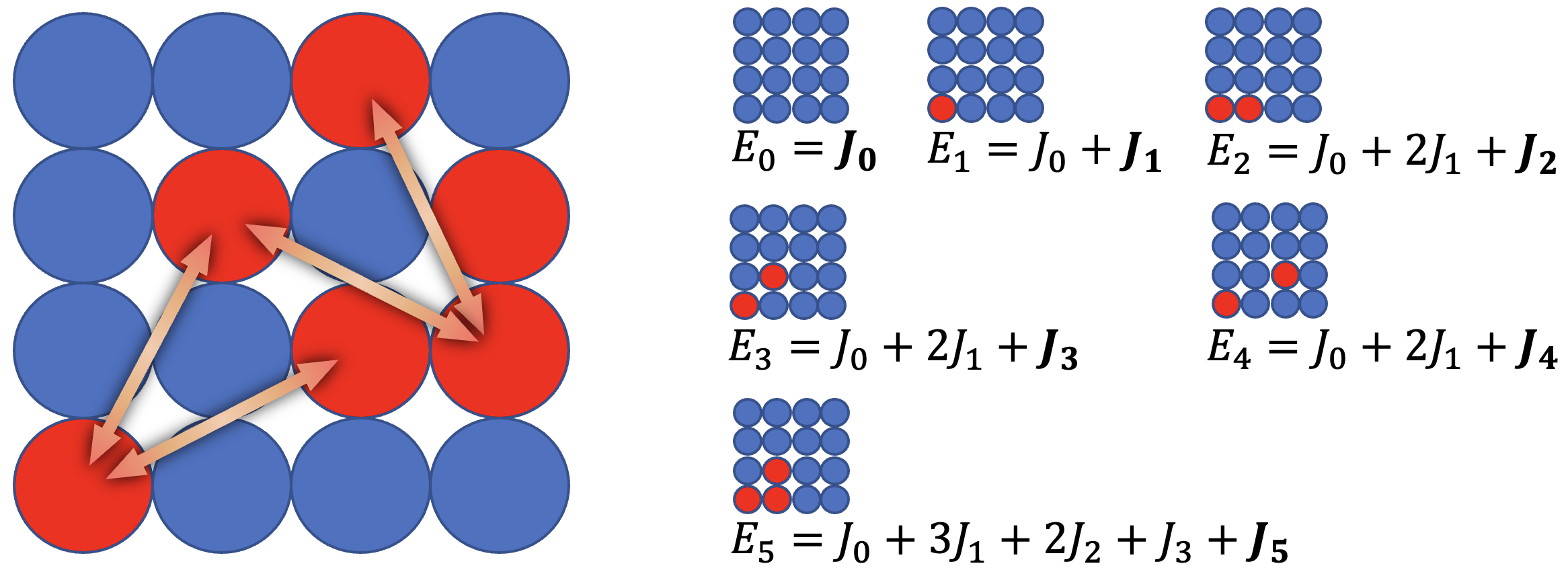}
\caption{Example of a two dimensional binary system. The properties (\eg the energy $E$) of an arbitrary configuration (left) can be approximately computed in terms of a superposition of effective \textit{n}-body interactions (right).}
\label{fig:basicsCE1}
\end{figure}

Figure \ref{fig:basicsCE1} shows a schema of a binary substitutional system where every site of a square lattice can be occupied by one of two species, indicated by blue and red color, respectively. The goal is to find the relation between the configuration of the substituent species (red) and the physical properties of the system, for instance the total energy. This question could be answered by performing numerically costly first-principles calculations for every atomic arrangement of the lattice. Though quite convenient, this simple strategy turns out to be impractical since the number of configurations scales as $2^N$, with $N$ being the number of lattice sites (for the system of Fig.~\ref{fig:basicsCE1}, the number of configurations would be of the order of $7\times10^4$, without considering symmetries). However, by making a few \textit{reasonable} assumptions, we can calculate the energy of an arbitrary configuration as in the left side of Fig.~\ref{fig:basicsCE1} in a very efficient way. The only input needed is the energies of a small number of well-chosen configurations. 

We start by considering \textit{effective} $n$-body interactions and assume that (i) the 1-body interactions are larger than the 2-body interactions, these being larger than the 3-body interactions, and so on; (ii) the interactions decrease with distance; and (iii) the total energy can be written as a linear superposition of \textit{n}-body interactions. Next, we evaluate the \textit{ab initio} energies $E_{0}$ to $E_{5}$ of the configurations depicted on the right side of Fig.~\ref{fig:basicsCE1}. According to the assumptions made above, these contain \textit{important} interactions that are labeled with $J_{0}$ to $J_{5}$: $J_{0}$ represents the energy $E_0$ of the pristine structure (all circles blue). $J_{1}$ accounts for the change in energy of the system when a "blue" species is replaced by a "red" one, and can be obtained by subtracting $J_{0}$ from the computed $E_1$. $J_{2}$ to $J_{4}$ represent 2-body interactions of increasing distance. They can be determined by making use of assumption (iii). For instance, the calculated value of $E_2$ is the addition of the energy of the pristine structure ($J_0$), plus two "blue"-to-"red" susbstitutions ($2 J_1$), plus an additional term ($J_2$) that embodies a 2-body interaction and can be estimated by the difference $E_2-(J_0+2J_1)$. Finally, $J_5$ represents a 3-body interaction that can be estimated analogous to the 2-body interactions. With these so-called effective cluster interactions $J_{i=0-5}$, we can now predict the energy of the configuration on the left side of Fig.~\ref{fig:basicsCE1} as 
\begin{equation}
\hat{E}=J_0+6J_1+2J_2+4J_3+4J_4+1J_5.
\label{eq:simple}
\end{equation} 
If assumptions (i) to (iii) are true, then the predicted energy value $\hat{E}$ will be close to the \textit{ab initio} energy $E$ (henceforth predicted values will be denoted by a "hat" symbol). The integer numbers multiplying the interactions $J_i$ tell how many times the corresponding pattern or \textit{cluster} is present in a given structure. For instance, it is 4 for the interaction $J_4$, as indicated by the four arrows in Fig.~\ref{fig:basicsCE1}. 

\begin{figure}[htpb]
\centering
\includegraphics[scale=0.55]{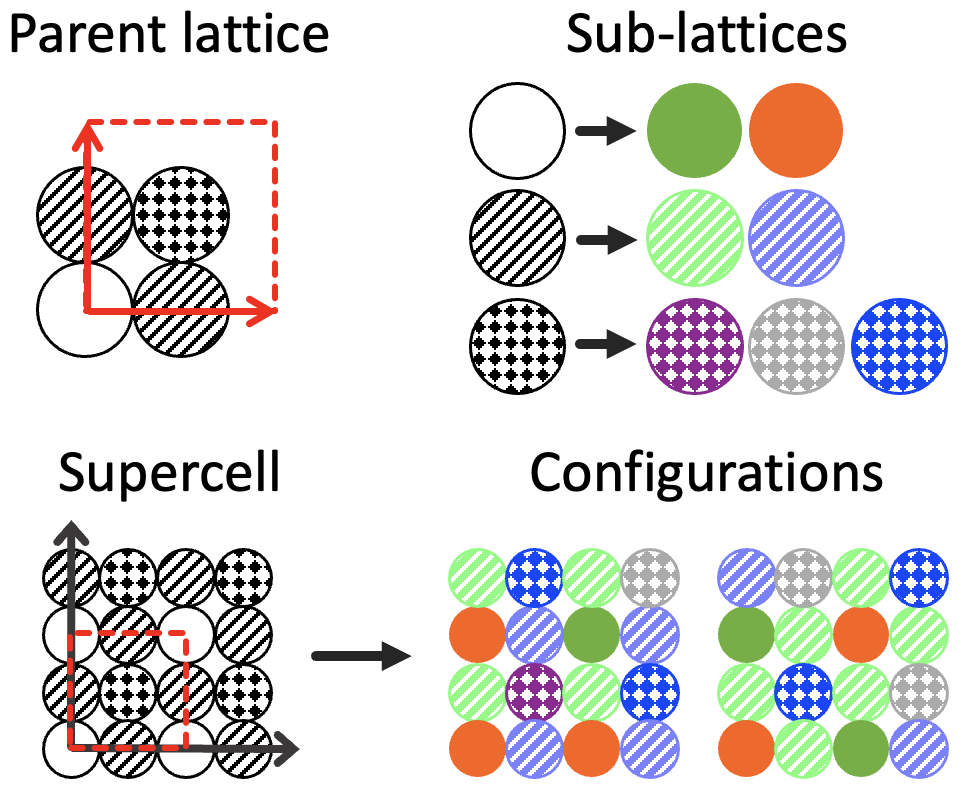}
\caption{Example of a two-dimensional system with a multi-composition and multi-sublattice setup. The parent lattice with four sites contains three different sub-lattices (top left). Each of them can host different numbers and kinds of atoms, as indicated by colors (top right). Supercells (bottom left) based on this parent lattice (dashed box), serve as "blue-print" for the construction of configurations (bottom right) compatible with the sublattice definition.}
\label{fig:mcml}
\end{figure}

The naive approach just exposed has two important shortcomings. First, the assumtions (i) and (ii) cannot be expected to be true for all material properties. Second, it cannot be easily adapted to more complex situations as, \eg shown in Fig.~\ref{fig:mcml}. In this case, we have a parent lattice (top left) with different \textit{sublattice types} assigned to the lattice points. These are indicated by white, striped, and gridded patterns. Different numbers of substitutions and species types can occupy each sublattice, as indicated by the colors (top right). Large supercells can be constructed by periodic repetitions of the parent lattice (bottom left). These allow for the construction of configurations compatible with the parent lattice, such as fully disordered, ordered, or partially (dis)ordered structures as shown in the bottom right. Analogous situations are frequently encountered, \eg in complex bulk or surface alloys.

In the following, we will explain how to treat this general case. The formulation does not make use of assumptions (i) and (ii) but allows for the application of AI methods to identify property-specifc interactions.  

\subsection{General CE formalism\label{ssec:general}}
We consider a crystal lattice with atomic positions $\mathbf{R}_i$, $i=1,...,N$. Every position $\mathbf{R}_i$ can host any of $M_i$ atoms of different type. An arbitrary arrangement of species in the lattice can be represented by vector $\bm{\sigma}=(\sigma_{1}, \sigma_{2}, ... , \sigma_{N})$, with $\sigma_{i}$ being integer numbers between $0$ and $M_i-1$, indicating the species at position $\mathbf{R}_i$. The pristine crystal is defined as the configuration with $\sigma_{i}=0$ for all $i$. 

A physical property that depends on the configuration can be represented by a function $P(\bm{\sigma})$. It can be shown~\cite{Sanchez19840,Walle2009} that, in the discrete space spanned by the vectors $\bm{\sigma}$, there exist complete and orthonormal sets of basis functions $\Gamma_{\bm{\alpha}}(\bm{\sigma})$, called \textit{cluster functions}, in terms of which $P(\bm{\sigma})$ can be expanded:
\begin{equation}
P({\bm{\sigma}}) = \sum_{\bm{\alpha}}J_{\bm{\alpha}}\Gamma_{\bm{\alpha}}(\bm{\sigma}).
\label{eq:ce}
\end{equation}
The real numbers $J_{\bm{\alpha}}$ are the expansion coefficients and $\bm{\alpha}$ stands for a vector of components  $\alpha_i\in\{0,1,....M_i-1\}$, $i=1,...,N$. 
%
The cluster functions fulfil the orthonormality condition $\langle \Gamma_{\bm{\alpha}},\Gamma_{\bm{\beta}}\rangle=\delta_{\bm{\alpha}\bm{\beta}}$. Here, $\delta_{\bm{\alpha}\bm{\beta}}=\prod_{i=1}^{N} \delta_{\alpha_i\beta_i}$, with $\delta_{\alpha_i\beta_i}$ being the Kronecker delta, and the inner product is defined by $\langle f,g\rangle=\sum_{\bm{\sigma}}f(\bm{\sigma})g(\bm{\sigma})/\prod_{i=1}^{N}M_i$, with $f$ and $g$ arbitrary functions in configuration space and the sum running over all possible configurations $\bm{\sigma}$ of the system. The cluster functions can be constructed as follows \cite{Sanchez19840, Walle2009}:
\begin{equation}
\Gamma_{\bm{\alpha}}(\bm{\sigma}) = \prod_{i=1}^N\gamma_{M_i,\alpha_i}(\sigma_i),
\label{eq:basis}
\end{equation}
with the functions $\gamma_{M_i,\alpha_i}(\sigma_i)$ forming a real, orthonormal basis in the discrete (and finite) domain $\sigma_i=0,\,1,...,\,(M_i-1)$. Usual choices for this basis include discrete Chebyshev polynomials and trigonometric functions~\cite{Sanchez19840,Walle2009,Wolverton1994}.  

Without loss of generality, one can choose the constant function $\gamma_{M,0}(\sigma)=1$ for $\alpha=0$. Thus, the product in Eq.~(\ref{eq:basis}) can be restricted to the indices $\alpha_i\ne0$. Accordingly, a cluster function is defined by indicating the set $\left\{ (i,\alpha_i) | \alpha_i\ne0\right\}$. The special case where all $\alpha_i$ are 0, corresponds to the \textit{empty cluster} function, $\Gamma_{\emptyset}(\bm{\sigma})=1$. There are $M_i$ basis functions $\gamma$ at a site $i$. Obviously, if two clusters $\bm{\alpha}$ and $\bm{\beta}$ are related by a symmetry operation of the parent lattice, then their expansion coefficients are equal, \ie $J_{\bm{\alpha}}=J_{\bm{\beta}}$. 
It should be noted that the sublattice types also determine the symmetry. Hence, the sum in Eq.~(\ref{eq:ce}) can be split into a sum over symmetrically inequivalent (s.i.) clusters and a sum over symmetrically equivalent ones:
\begin{equation}
P({\bm{\sigma}})
= \sum_{\bm{\alpha}}^{\text{s.i.}} {\cal M}_{\bm{\alpha}}  J_{\bm{\alpha}} X_{\bm{\alpha}}({\bm{\sigma}}).
\label{eq:sumsplit}
\end{equation}
Here, the \textit{cluster correlation function}  $X_{\bm{\alpha}}({\bm{\sigma}})$ 
\begin{equation}
X_{\bm{\alpha}}(\bm{\sigma})=\frac{1}{{\cal M}_{\bm{\alpha}}}\sum_{\bm{\beta} \in {\cal{O}}(\bm{\alpha})}\Gamma_{\bm{\beta}}(\bm{\sigma}).
\label{eq:correlations}
\end{equation}
is the average of the cluster functions in the set of clusters symmetrically equivalent to $\bm{\alpha}$. This set, of size ${\cal M}_{\bm{\alpha}}$, is called the \textit{orbit} of cluster $\bm{\alpha}$, and we denote it by ${\cal{O}}{(\bm{\alpha})}$.

Models are often constructed for an intensive property, as for instance the energy per unit cell. In such a case,  ${\cal M}_{\bm\alpha}$ in Eq.~(\ref{eq:sumsplit}) can be replaced by $m_{\bm\alpha} = {\cal M}_{\bm\alpha}V_{pc}/V_{sc}$, with $V_{sc}$  and $V_{pc}$ being the super- and the parent-cell volume, respectively. The integers $m_{\bm\alpha}$ are also called cluster multiplicities. In the simple binary case of Fig.~\ref{fig:basicsCE1}, we have $M_i=2$. If we choose the basis $\gamma_{20}(\sigma)=1$,  $\gamma_{21}(\sigma)=\sigma$, $\sigma=0,1$ \footnote{This basis is not orthonormal, however, this is not problematic for the present example.}, it is easy to verify that, for the structure represented on the left side of Fig.~\ref{fig:basicsCE1}, the cluster correlations for clusters 1 to 5 on the right are, respectively, $X=6/16$, $2/32$, $4/32$, $4/64$, and $1/64$, with the denominators being the corresponding values of ${\cal M}$. Thus, the values of ${\cal M}X=6,\, 2,\, 4,\, 4,\, 1$ agree with the coefficients for $J_1$  to $J_5$ in Eq.~(\ref{eq:simple}).

The number of clusters in the expansion of Eq.~(\ref{eq:sumsplit}) is in principle infinite, and the expansion coefficients $J_{\bm{\alpha}}$ (called effective cluster interactions, ECIs in short), are still undetermined. For practical applications, though, it is necessary to cut off the cluster basis, keeping only the most \textit{relevant} clusters in Eq.~(\ref{eq:sumsplit}). One also has to determine the ECIs that lead to accurate predictions of the property of interest. These are the main tasks to be addressed in the construction of CE models. Below, we briefly explain how this problem is tackled by using AI techniques. 

\subsection{CE as a data-analytics problem\label{ssec:ceda}}
To build a CE model of a material property of interest, we need (i) a set of structures ${\cal{S}}=\{\bm{\sigma}_1, \bm{\sigma}_2,...,\bm{\sigma}_{N_s}\}$, (ii) the corresponding calculated properties $\bm{P}^\top=(P_1, P_2, ..., P_{N_s})$ with $P_i=P(\bm{\sigma}_i)$, and (iii) a set of clusters ${\cal{C}}=\{\bm{\alpha}_1, \bm{\alpha}_2, ...,\bm{\alpha}_{N_c}\}$. Then, we build a matrix $\bm{X}$ of cluster correlations, with elements $X_{ij}=X_{\bm{\alpha}_j}(\bm{\sigma}_i)$, such that a column $j$ in the matrix represents a cluster, while a row $i$ represents a structure. By using Eq.~(\ref{eq:sumsplit}) with the sum on clusters limited to the set $\cal{C}$, we can write, for an intensive property
\begin{equation}
\hat{\bm{P}}=\bm{X}\bm{{\cal J}} ,
\label{eq:linear_model}
\end{equation}
where $\bm{{\cal J}}$ is a column vector with elements ${\cal{J}}_i=m_{\bm{\alpha}_i}{{J}}_{\bm{\alpha}_i}$, $i=1,...,N_c$. Optimal cluster interactions $\bm{{\cal{J}}}$ can be found by minimizing a cost function:
\begin{equation}
\bm{{\cal J}} = 
\underset{\bm{{\cal J}}^*}{\mbox{argmin}}
\left[\lVert{\bm{X}\bm{{\cal J}}^*-\bm{P}}\rVert_2^2+\phi(\bm{{\cal J}}^*)\right].
\label{eq:mse}
\end{equation}
Here, we use the standard definition of the $\ell_p$-norm of a vector $\bm{x}=(x_1,...,x_n)$, namely $\lVert\bm{x}\rVert_p=(\sum_{i=1}^n |x_i|^p)^{1/p}$. (In the particular case, $p=0$, $\lVert\bm{x}\rVert_0$ is defined as the number of non-zero components of $\bm{x}$.) The first term inside the square brackets is the mean squared error (MSE) of the predictions, and the second term, $\phi(\bm{{\cal J}}^*)$, is a penalization or regularization term. The latter can be used for different purposes. For instance, if there are no linearly dependent columns in $\bm{X}$ and  $N_c\le N_s$, \ie the number of clusters is smaller than the number of structures in the training set ${\cal{S}}$, optimal cluster interactions $\bm{{\cal{J}}}$ can found by directly minimizing the MSE of the predictions ($\phi=0$), with the solution $\bm{{\cal J}} = \left(\bm{X}^\top\bm{X}\right)^{-1}\bm{X}^\top\bm{P}$.

In typical applications, data are scarce, and one wants to find the relevant interactions out of a large set of clusters, \ie solve an underdetermined problem with $N_c> N_s$. In this case, the Gram matrix $\bm{X}^\top\bm{X}$ cannot be inverted. Thus, the optimization problem posed by Eq.~(\ref{eq:mse}) must be regularized. The easiest way to achieve this is by choosing $\phi( \bm{{\cal J}})=\lambda \lVert \bm{{\cal J}}\rVert_2^2$, with the hyperparameter $\lambda\in\Re$. In this case, the solution is $\bm{{\cal J}} = \left(\bm{X}^\top\bm{X}+\lambda \bm{I}\right)^{-1}\bm{X}^\top\bm{P}$ \cite{Deisenroth2020}. Besides making the problem solvable, this selection penalizes large ECI values, leading to solutions with increasingly small interactions for increasing $\lambda$. To obtain sparse models in which the relevant interactions are represented by a small number of clusters, compressed-sensing techniques \cite{Foucart2013,Nelson2013} may be employed. In compressed sensing, one seeks penalization terms in Eq.~(\ref{eq:mse}) that promote sparsity,  \ie giving solutions with many interactions ${\cal{J}}_i$ being exactly zero. Common penalization terms fulfilling this requirement, are $\phi=\lambda \lVert \bm{{\cal J}}\rVert_0$ or $\phi=\lambda \lVert \bm{{\cal J}}\rVert_1$. The first choice penalizes solutions with a large number of non-zero parameters by using the $\ell_0$-norm, thus producing sparse models. The associated minimization problem is NP-hard \cite{Arora2009},  \ie its solution cannot be found in polynomial time, and exact solutions can only be computed for rather small clusters pools. Using the second choice (also called the {\it Manhattan norm}), leads to the least-absolute-shrinkage-and-selection-operator (LASSO) approach \cite{Tibshirani1996}, that represents a convex optimiztion problem, and efficient algorithms exist for its solution. Under certain conditions \cite{Ghiringhelli2017}, the solutions found with LASSO may approximate well the solutions found using the $\ell_0$ norm.

Once the optimal $\bm{{\cal J}}$ is found by solving Eq.~(\ref{eq:mse}), one can use Eq.~(\ref{eq:sumsplit}) to perform property predictions for arbitrary configurations $\bm{\sigma}$.

\section{Cluster expansion with \texttt{CELL}\label{sec:ce}}

\begin{figure}[htpb]
\centering
\includegraphics[scale=1.0]{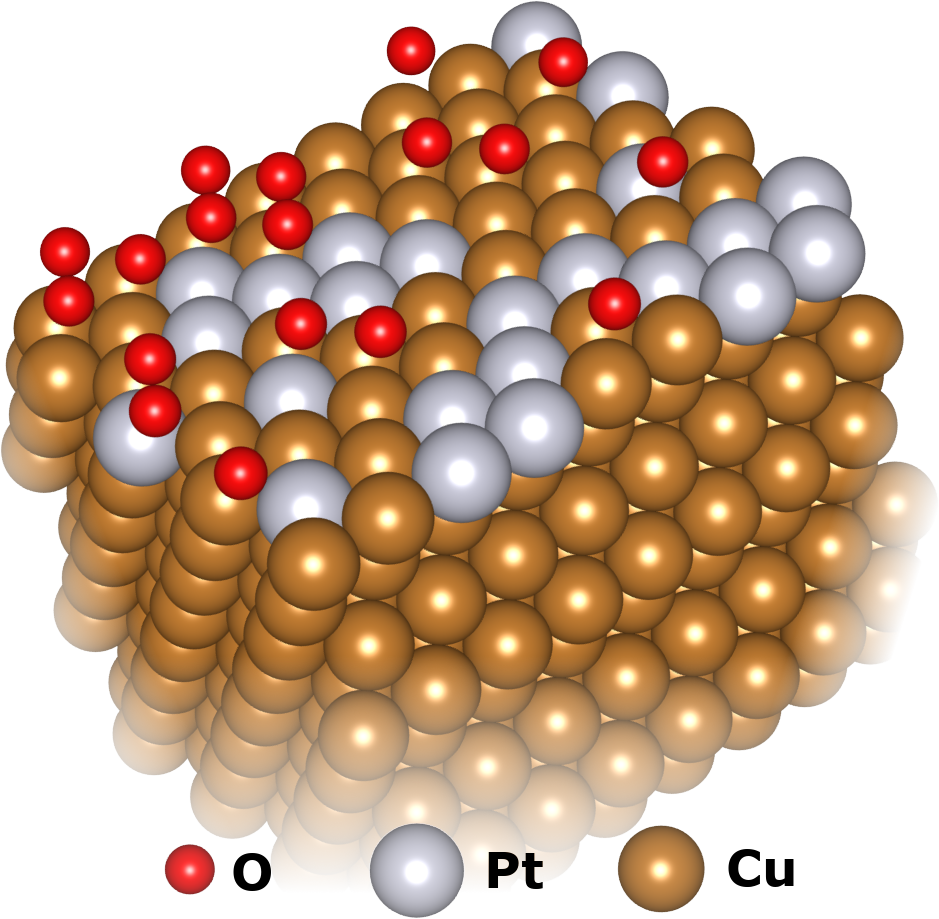}
\caption{Pt/Cu(111) surface alloy with atomic O adsorption.}
\label{fig:optcu}
\end{figure}

To demonstrate the construction of CE models with \texttt{CELL}, we build one for the energy of formation of a complex surface system, consisting of a Pt/Cu(111) surface alloy with atomic O adsorption, as shown in Fig.~\ref{fig:optcu}. In this example, both O adsorption and Pt-Cu alloying phenomena are simultaneously accounted for. Platinum and copper form a two-dimensional (2D) alloy in the top-most atomic layer of Cu(111), creating disordered as well as ordered 2D surface patterns (\eg p(2$\times$2) and $\sqrt{3}\times\sqrt{3}~R~30^\circ$ reconstructions) \cite{Lucci2014}. Atomic oxygen adsorbs preferentially on hollow fcc sites, both on the pristine Cu(111) \cite{Frey2014} and Pt(111) \cite{Frey2014,Schmidt2012} surfaces. These facts define the parent lattice of our system.

To avoid the generation of numerically costly DFT data sets, the total energies in the present example are calculated with the effective medium theory (EMT) calculator from ASE \cite{Larsen2017}. While the EMT potentials for Pt and Cu are quite realistic \cite{Jacobsen1996}, this is not the case for oxygen. Since, however, the main purpose of this section is to showcase the construction of a CE model and its use and not to capture the actual physical details of the system, we proceed in this way. The use of toy total-energy models is a common practice to test CE methods (see \eg Refs.~\cite{Nelson2013,Herder2015,Nguyen2017}).

All the code listings shown in this section are written in Python 3 \cite{Van-Rossum2009} and can be run interactively as Jupyter notebooks \cite{Kluyver2016} (see Sec.~\ref{sec:avail}). 

\subsection{Generation of structures}
The basic building block for the generation of structures for the CE is the parent lattice. It comprises the definition of the primitive cell of the pristine crystal and the species that can possibly occupy any crystal site. In \texttt{CELL}, a parent lattice is embodied by the \texttt{ParentLattice} class. It admits the definition of a multi-sublattice and multi-composition framework, as demonstrated in Listing~\ref{lst:p-lat}. In lines 1 to 5 of the listing, the Atomic Simulation Environment (ASE) \cite{Larsen2017} is used to create an \texttt{Atoms} object representing the primitive cell of an fcc~(111) Cu slab with three atomic layers and vacancy sites on hollow fcc positions (vacancies are indicated with the character \texttt{X}). This \texttt{Atoms} object is assigned to the variable \texttt{pristine}, as it represents the pristine primitive cell, hosting no substituents. In line 8, the \texttt{ParentLattice} class of \texttt{CELL} is loaded \footnote{Note that the \texttt{CELL} package is called and imported in Python code with the name \texttt{clusterx}.}, and a \texttt{ParentLattice} object is initialized in line 11 and assigned to the variable \texttt{p\_lat}. The initialization takes two arguments: the primitive cell (variable \texttt{pristine}), and a list of possible species that every crystal site can host. The latter is defined in the list \texttt{symbols} in line 10 of the listing: The first two layers may only contain Cu (thus behaving as \textit{spectator} atoms \cite{Walle2002},  \ie they determine the symmetry but are not part of any cluster), while the Cu atoms on the top layer may be substituted by Pt, and the remaining vacancy sites (X) can be substituted by oxygen. Note that the order of this list must correspond to the atomic arrangement in the \texttt{Atoms} object \texttt{pristine}. The latter could be inquired with the method \texttt{pristine.get\_positions()}.

\noindent
\begin{minipage}{\linewidth}
\begin{lstlisting}[caption={Creation of a two-dimensional multi-composition multi-sublattice \texttt{ParentLattice} object.}, label={lst:p-lat}]
from ase.build import fcc111, add_adsorbate

pristine = fcc111('Cu', a=3.59, size=(1,1,3)) # 3-atomic-layer Cu(111) slab
add_adsorbate(pristine,'X',1.7,position='fcc') # Hollow fcc vacancy site
pristine.center(vacuum=10.0, axis=2) # add vacuum along z-axis

# Note: CELL is imported with the name "clusterx"
from clusterx.parent_lattice import ParentLattice 

symbols = [['Cu'],['Cu'],['Cu','Pt'],['X','O']]
p_lat = ParentLattice(pristine, symbols=symbols)
p_lat.print_sublattice_types()
\end{lstlisting}
\end{minipage}
Upon execution of this code, the output, as shown in Fig.~\ref{fig:sublatt}, is displayed, indicating that three sublattices were created: two different binary sublattices (assigned sublattice types 0 and 2) and one spectator sublattice (sublattice type 1). In this way, a full multi-component multi-sublattice framework can be generated. \texttt{CELL} does not restrict the number of n-ary (\ie unary, binary, ternary, ...) sublattices that can be defined in this way.
\begin{figure}[htpb]
\centering
\begin{BVerbatim}[fontsize=\small]
+----------------------------------------------------------------+
|            The structure consists of 3 sublattices             |
+----------------------------------------------------------------+
| Sublattice type |     Chemical symbols     |  Atomic numbers   |
+----------------------------------------------------------------+
|        0        |        ['X' 'O']         |       [0 8]       |
|        1        |          ['Cu']          |       [29]        |
|        2        |       ['Cu' 'Pt']        |      [29 78]      |
+----------------------------------------------------------------+
\end{BVerbatim}
\caption{Output of Listing~\ref{lst:p-lat}, corresponding to a \texttt{ParentLattice} object with three sublattices.}
\label{fig:sublatt}
\end{figure}

\begin{figure}[htpb]
\centering
\includegraphics[width=.8\textwidth]{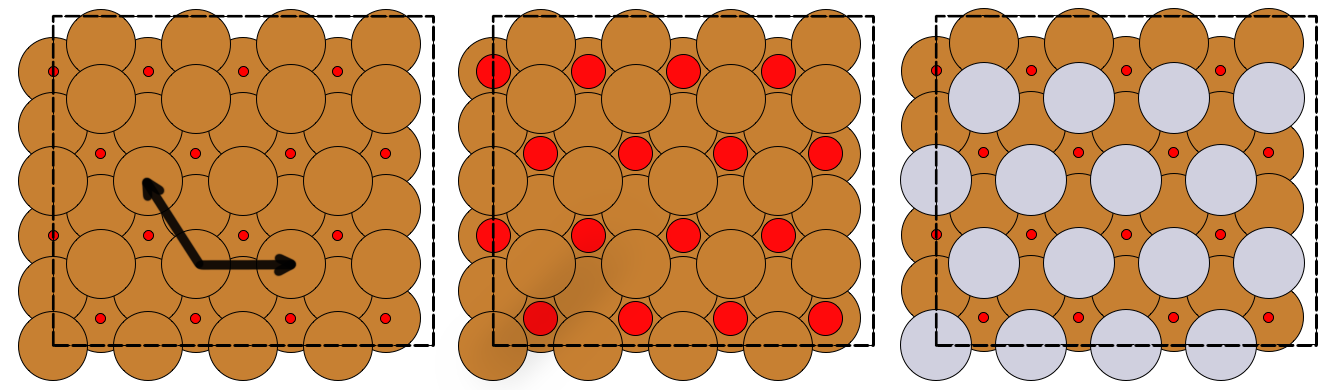}
\caption{ Graphical represention of a \texttt{SuperCell} object. Brown, small red, big red, and gray spheres represent copper, vacancies, oxygen, and platiunum atoms, respectively. From left to right: pristine supercell; full substitution of vacant hollow fcc sites by O atoms; full substitution of the top Cu layer by Pt atoms. Arrows indicate the unit cell vectors of the underlying parent lattice.}
\label{fig:scell}
\end{figure}

We can now build a supercell and visualize it, as shown in Listing~\ref{lst:scell}. In  line 2 of the listing, we use the \texttt{SuperCell} class of \texttt{CELL} to create a  \texttt{SuperCell} object, based on the parent lattice created in Listing~\ref{lst:p-lat}. The lattice coordinates of the supercell vectors are (4,0) and (-2,4), referring to the hexagonal unit vectors of the Cu(111) surface (indicated by arrows in the left panel of Fig.~\ref{fig:scell}). The call to the \texttt{juview} method (\texttt{CELL}'s plotting interface for Jupyter notebooks) in line 5 produces the image shown in Fig.~\ref{fig:scell}. In this graphical representation of the \texttt{SuperCell}, the first image on the left depicts the pristine, non-substituted crystal, while the images on the right, represent the results of substituting one of the species as allowed in the definition of the parent lattice.
\begin{minipage}{\linewidth}
\begin{lstlisting}[caption={Creation and visualization of a \texttt{SuperCell} object.}, label={lst:scell}]
from clusterx.super_cell import SuperCell
scell = SuperCell(p_lat,[[4,0],[-2,4]])

from clusterx.visualization import juview
juview(scell)
\end{lstlisting}
\end{minipage}

A supercell like the one depicted in Fig.~\ref{fig:scell}, serves as a blueprint for the generation of structures. In Listing~\ref{lst:sset}, we use the supercell to construct random structures and gather them in a \texttt{StructuresSet} object, which is created in line 2 of the listing. Objects of this class act as structure containers which have a database attached. They can be serialized, \eg in the form of JSON database files fully compatible with ASE's JSON databases.

\begin{minipage}{\linewidth}
\begin{lstlisting}[caption={Creation of a \texttt{StructuresSet} object containing 50 random structures.}, label={lst:sset}]
from clusterx.structures_set import StructuresSet
sset = StructuresSet(p_lat)
for i in range(50):
	rnd_str = scell.gen_random_structure()
	sset.add_structure(rnd_str)
juview(sset,n=4)
\end{lstlisting}
\end{minipage}
In line 4, structure creation takes place by calling the \texttt{gen\_random\_structure()} method of the \texttt{SuperCell} class. This returns a \texttt{Structure} object. In \texttt{CELL}, \texttt{Structure} objects consist of a supercell augmented with a \textit{decoration} array, which is a list indicating which species occupy the individual sites of the supercell. Thus, all relevant information like, \eg sublattice types, is provided. Less memory-consuming representations consisting of only the decoration array, are available for tasks such as Monte Carlo simulations (see Sec.~\ref{sec:mc}). 

\begin{figure}[htpb]
\centering
\includegraphics[width=0.8\textwidth]{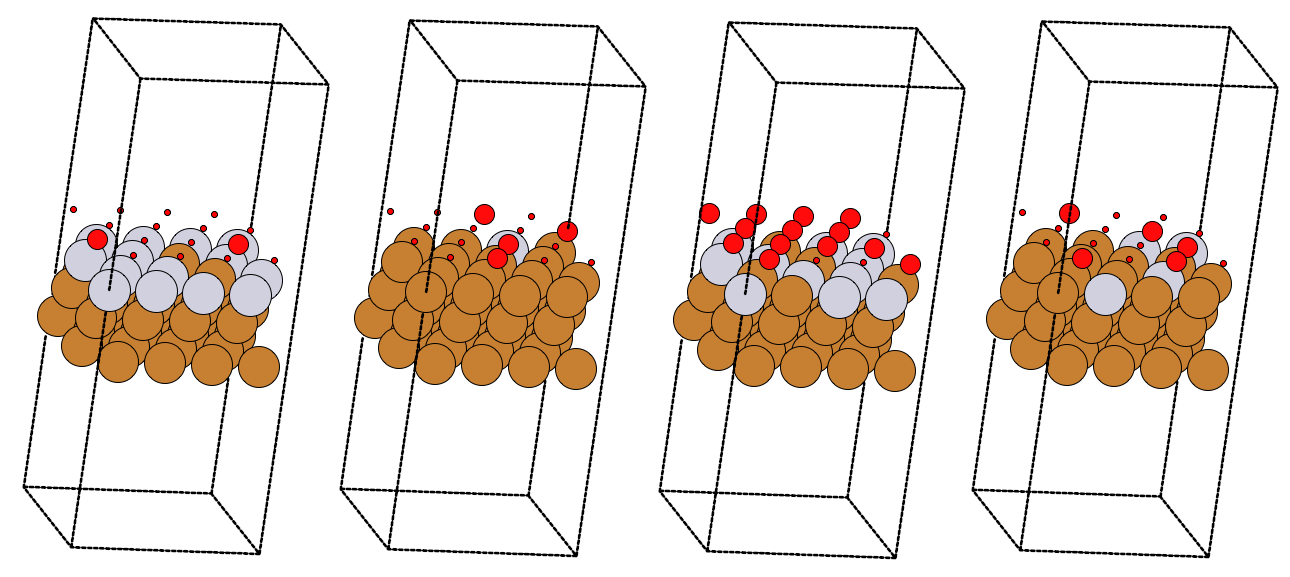}
\caption{Four simulation cells of random structures belonging to the \texttt{StructuresSet} object of Listing~\ref{lst:sset}. The color code is the same as in Fig.~\ref{fig:scell}.}
\label{fig:sset}
\end{figure}

The 50 generated structures are added to \texttt{sset} in line 5 of Listing~\ref{lst:sset} by using the \texttt{add\_structure()} method of the \texttt{StructuresSet} class. Finally, four of the generated structures are displayed by calling the \texttt{juview()} method in line 6, with the result as shown in Fig.~\ref{fig:sset}. 

\begin{figure}[htpb]
\centering
\includegraphics[scale=0.45]{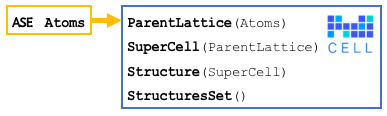}
\caption{Inheritance map of \texttt{CELL}'s classes for structure construction and collection. Bold-face words indicate class names, and parenthesis indicate inheritance. Yellow box: \texttt{Atoms} class from ASE, blue box: classes belonging to \texttt{CELL}. }
\label{fig:structure}
\end{figure}
In summary, the construction and collection of structures in \texttt{CELL} takes place in four classes, three of them related by inheritance as shown in Fig.~\ref{fig:structure}: The \texttt{ParentLattice} class inherits from ASE's \texttt{Atoms} class and is supplied with a multi-composition-multi-sublattice framework; a \texttt{SuperCell} is an enlarged \texttt{ParentLattice}; a \texttt{Structure} is a \texttt{Supercell} \textit{decorated} with a specific configuration of substituent species. These classes are equipped with a lot of useful functionality, either through methods inherited from ASE's \texttt{Atoms} class, or from methods native to \texttt{CELL}. The latter are documented in Ref. \cite{Cell2019}. \texttt{StructuresSet} objects allow, \eg for the union of sets through the "+" operator, serialization, aggregation of properties, and more.

\subsection{Calculation of configuration-dependent properties}
Now that we have a set of structures, the next step in a standard CE workflow is to perform \textit{ab initio} computations of the properties that we want to predict with a CE model. A common property to be modeled is the total internal energy of the alloy, $E(\bm{\sigma})$. With \texttt{CELL}, its computation can be easily achieved through the method \texttt{calculate\_property()} of the \texttt{StructuresSet} class. All \textit{ab initio} codes supported by the calculators of ASE can be employed. This is exemplified in the following listing by using the effective-medium-theory (\texttt{EMT}) calculator:
\begin{minipage}{\linewidth}
\begin{lstlisting}[caption={Calculation of the total energy of every structure in a \texttt{StructuresSet} object with a calculator of ASE.}, label={lst:calcprop1}]
from ase.calculators.emt import EMT 
sset.set_calculator(EMT()) # Assign EMT() calculator to StructuresSet.

# The total energy of every structure in sset is evaluated.
sset.calculate_property(prop_name="e_tot") 
\end{lstlisting}
\end{minipage}
Here, in line 1, the \texttt{EMT} calculator is loaded and an instance of it is attached to the \texttt{StructuresSet} object in line 2. Finally, in line 5 the calculator is used to compute the total energy of every structure in the set. By using the \texttt{prop\_name} argument, we assign the name \texttt{"e\_tot"} to this property. This acts, one the one hand, as a label to later retrieve the property values and, on the other hand, for storage upon serialization in, \eg a JSON file, with the \texttt{serialize()} method of \texttt{StructuresSet}.

For a surface system as the one we study here, a physically meaningful quantity is the adsorption energy, $E_{\text{ads}}$, defined as:
\begin{equation}
E_{\text{ads}}(\bm{\sigma})=\frac{1}{N}\left[E(\bm{\sigma})-n_{\text{O}} \frac{1}{2}E_{\text{O}_2}-n_{\text{Pt}}(E_{\text{Pt,bulk}}-E_{\text{Cu,bulk}})-E_{\text{Cu,surf.}}\right]
\label{eq:eads}
\end{equation}
Here, $N$ is the number of Cu-Pt sites in the top-most layer of the supercell, $n_\text{O}$ ($n_\text{Pt}$) the number of O (Pt) atoms, $E_{\text{O}_2}$ the energy of an O$_2$ molecule, $E_{\text{Pt (Cu),bulk}}$ the energy per atom of fcc Pt (Cu), and $E_{\text{Cu,surf.}}$ the total energy of the pristine Cu slab. Since structural relaxation can notably affect the relative stability of structures, it is important to build the CE model with energies $E(\bm{\sigma})$ corresponding to optimized structures. The required steps are implemented in a custom python function, \texttt{ads\_energy} (see Sec.~\ref{sec:avail}), which gets a \texttt{Structure} object as argument and returns $E_{\text{ads}}$. Finally, the method \texttt{calculate\_property} is called as follows:

\begin{minipage}{\linewidth}
\begin{lstlisting}[caption={Calculation of the absorption energy of every structure in a \texttt{StructuresSet} object by means of the custom function \texttt{ads\_energy}.}, label={lst:e_ads}]
sset.calculate_property(prop_name="e_ads", prop_func=ads_energy) 
\end{lstlisting}
\end{minipage}
In more detail, \texttt{ads\_energy} performs a structure optimization using the BFGS method. Three constraints are applied here: (i) The bottom-most Cu layer is fixed, (ii) the top-most Pt-Cu layer is allowed to relax in the $(x,y)$ plane only, (iii) the O positions may relax in the perpendicular direction, $z$ only. With the latter two constraints, we avoid reconstructions that may become significant at large Pt concentrations (\eg a Pt atom moving out from the Pt-Cu layer, or triads of O atoms sitting at bridge positions). We do so here for the sake of simplicity, although the degree of complexity brought about by these reconstructions could still be accounted for with \texttt{CELL} by adding the corresponding sites in the parent-lattice definition. 

After evaluating the energies, we can plot the result by using the \texttt{plot\_property\_vs\_concentration} method from \texttt{CELL}'s visualization package:

\noindent
\begin{minipage}{\linewidth}
\begin{lstlisting}[caption={Code for generating a plot of the \textit{ab initio} adsorption energy as a function of the Pt concentration.},label={lst:visua1}]
from clusterx.visualization import plot_property_vs_concentration
plot_property_vs_concentration(sset, site_type=2, property_name ="e_ads")
\end{lstlisting}
\end{minipage}

\begin{figure}[htpb]
\centering
\includegraphics[]{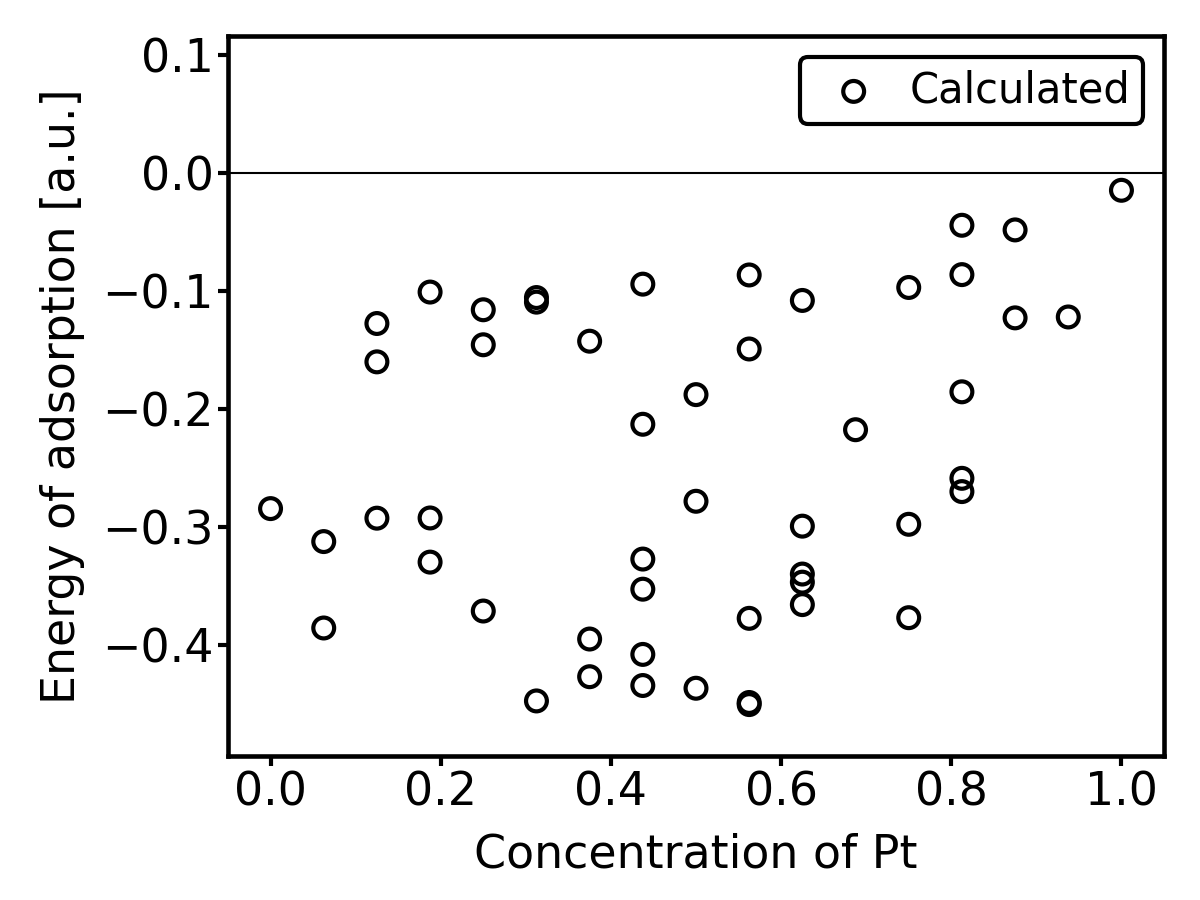}
\caption{Energy of adsorption of 50 optimized random structures. The plot was created with Listing~\ref{lst:visua1}.}
\label{fig:optcu-rnd}
\end{figure}

Here, we instruct \texttt{CELL} to plot the adsorption energy calculated before (labeled \texttt{e\_ads}) versus the concentration of the type-2 sublattice (\ie the Pt concentration in the Cu-Pt surface alloy). The result is shown in Fig.~\ref{fig:optcu-rnd}.

In this case, we have exemplified the calculation of properties by either using ASE's calculators (Listing~\ref{lst:calcprop1}) or custom functions provided by the user (Listing~\ref{lst:e_ads}). An approach consisting of the creation of folders containing input files for \textit{ab initio} packages is also possible \cite{Cell2019}.

\subsection{Clusters pool}

\begin{figure}[htpb]
\centering
\includegraphics[width=0.5\textwidth]{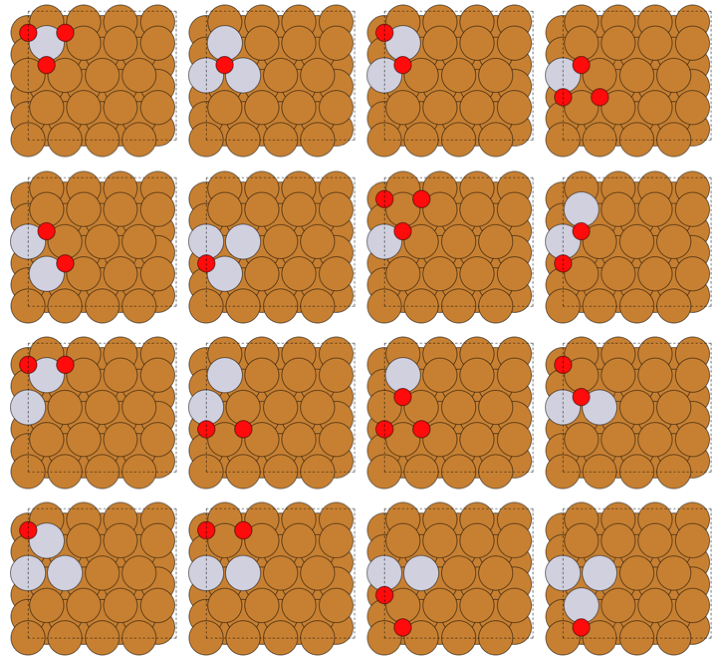}
\caption{4-point clusters with radii less than 4.3 \AA\ gererated by \CELL. The color code is not to be associated to atom types but to the index $\alpha_i$ of the basis functions $\gamma_{M_i,\alpha_i}(\sigma_i)$ of Eq.~(\ref{eq:basis}):  For the top-layer sites (sublattice type 2, see Fig.~\ref{fig:sublatt}), brown and gray correspond to $\alpha=0$ and $1$, respectively. For the fcc hollow sites (sublattice type 0), red circles correspond to $\alpha=1$, while the remaining ones (not indicated) correspond to $\alpha=0$.}
\label{fig:cloptcu}
\end{figure}

Having the training data constructed, we have to define which cluster functions will constitute our basis set. This is done with the \texttt{ClustersPool} class of \texttt{CELL}, as shown in the following listing:

\noindent
\begin{minipage}{\linewidth}
\begin{lstlisting}[caption={Generation of a pool of clusters.},label={lst:codepool}]
from clusterx.clusters.clusters_pool import ClustersPool
cpool = ClustersPool(p_lat, 
				npoints=[1,2,3,4], 
				radii=[0,-1,-1,4.3], 
				super_cell=scell)
				
cpool.serialize(db_name="cpool.json")
cpool.print_info()
\end{lstlisting}
\end{minipage}
\noindent
In line 1, we start by importing the \texttt{ClustersPool} class of \texttt{CELL}. Then, in lines 2-5, an instance of \texttt{ClustersPool} is created and assigned to the variable \texttt{cpool}. Clusters with 1 to 4 points are created, as indicated with the argument \texttt{npoints} in line 3. In line 4, the \textit{radius} of every cluster, \ie the maximum distance between any of its sites, is specified. This is obviously 0 for the 1-point cluster. For clusters with 2 and 3 points, a negative radius (-1) is assigned. A negative number is used to indicate that all unique clusters compatible with the periodic boundary conditions of the supercell \texttt{scell}, specified in line 5 by the argument \texttt{super\_cell}, are generated. For clusters with 4 points, we indicate a small radius of 4.3 \AA. This selection is motivated by the notion that clusters with many points should mainly capture short-range-order effects. No general rule exists, nevertheless, and one should try different parameters for other properties or systems. The execution of line 7 serializes the \texttt{ClustersPool} to a JSON file. The created pool of clusters can then be conveniently visualized, for instance through the graphical user interface of ASE. Also using the \texttt{juview} function of \texttt{CELL} is possible, as explained before for the visualization of the \texttt{StructuresSet} (see Listing~\ref{lst:sset} and Fig.~\ref{fig:sset}). In Fig.~\ref{fig:cloptcu}, all the generated 4-point clusters are shown. The execution of the last line generates the output shown in Fig.~\ref{fig:cpoolinfo}, listing index, number of points, radius, and multiplicity of every generated cluster.

\begin{figure}[htpb]
\centering
\begin{BVerbatim}[fontsize=\small]
+-----------------------------------------------------------------------+
|                         Clusters Pool Info                            |
+-----------------------------------------------------------------------+
|      Index      |  Nr. of points  |     Radius      |  Multiplicity   |
+-----------------------------------------------------------------------+
|        0        |        1        |      0.000      |        1        |
|        1        |        1        |      0.000      |        1        |
|        2        |        2        |      2.245      |        3        |
        ...               ...               ...               ... 
|       121       |        4        |      4.234      |        3        |
|       122       |        4        |      4.234      |        3        |
+-----------------------------------------------------------------------+
\end{BVerbatim}
\caption{Output of Listing~\ref{lst:codepool}.}
\label{fig:cpoolinfo}
\end{figure}

\subsection{Building CE models \label{ssec:build}}
The essential elements for the construction of a cluster expansion model are now available, namely, the training data contained in the \texttt{StructuresSet} object \texttt{sset} and the \texttt{ClustersPool} object \texttt{cpool}. With them, we can obtain the vector $\bm{P}$ and the input matrix $\bm{X}$ entering Eq.~(\ref{eq:mse}), as demonstrated in Listing~\ref{lst:xandp}.

\begin{figure}[htpb]
\begin{lstlisting}[caption={Code to set up the input matrix $\bm{X}$ (variable \texttt{x} in line 5) and the vector of calculated properties $\bm{P}$ (variable \texttt{p} in line 8) of Eq.~(\ref{eq:mse}).}, label={lst:xandp}]
from clusterx.correlations import CorrelationsCalculator
corrcal = CorrelationsCalculator("trigonometric", p_lat, cpool)

# Compute full correlations matrix for sset
x = corrcal.get_correlation_matrix(sset)

# Extract ab initio property values
p = sset.get_property_values("e_ads")
\end{lstlisting}
\end{figure}

Here, we start by creating a \texttt{CorrelationsCalculator} object in line 2. With this object, which is assigned to the variable \texttt{corrcal}, we can calculate the cluster correlations of Eq.~(\ref{eq:correlations}) for any structure based on the parent lattice \texttt{p\_lat} and for all clusters in \texttt{cpool}. In this case, the calculator uses basis functions $\gamma_{M,\alpha}(\sigma)$ based on trigonometric functions, as indicated by the argument \texttt{"trigonometric"}. The full matrix $\bm{X}$ and the vector $\bm{P}$  are created in lines 5 and 8, respectively.

We now build a CE model by using the \texttt{EstimatorFactory} class of \texttt{CELL}. This class acts as an interface to all the estimators of the \texttt{scikit-learn} machine-learning Python library \cite{Pedregosa2011,Buitinck2013} and to \texttt{CELL}'s native estimators. 

\begin{figure}[htpb]
\begin{lstlisting}[caption={Creation of ridge-regression model with the \texttt{EstimatorFactory} and \texttt{Model} objects of \texttt{CELL}.}, label={lst:ridge}]
from clusterx.estimators.estimator_factory import EstimatorFactory
from clusterx.model import Model

# Build and fit a ridge-regression estimator of scikit-learn
re = EstimatorFactory.create("skl_Ridge", alpha=1e-8) 
re.fit(x,p)
ce_model = Model(corrcal, "e_ads", estimator = re) # Create a CE model
ce_model.report_errors(sset)
\end{lstlisting}
\end{figure}

In line 5 of Listing~\ref{lst:ridge}, a ridge-regression estimator from \texttt{scikit-learn} is created. This estimator solves Eq.~(\ref{eq:mse}) with an $\ell_2$ regularization. The value \texttt{alpha=1e-8} indicates a small regularization parameter$\lambda=10^{-8}$. Finally, a CE model is created and assigned to the variable \texttt{ce\_model} (line 7), and the errors are reported (line 8).
\begin{figure}[htpb]
\centering
\begin{BVerbatim}[fontsize=\small]
+-----------------------------------------------------------+
|                Report of Fit and CV scores                |
+-----------------------------------------------------------+
|                   |        Fit        |        CV         |
+-----------------------------------------------------------+
|       RMSE        |      0.00000      |      0.01412      |
|        MAE        |      0.00000      |      0.01146      |
|       MaxAE       |      0.00000      |      0.03907      |
+-----------------------------------------------------------+
\end{BVerbatim}
\caption{Output of Listing~\ref{lst:ridge}.}
\label{fig:outputridge}
\end{figure}
The output, presented in Fig.~\ref{fig:outputridge}, shows that the root mean square error (\texttt{RMSE}), the mean absolute error (\texttt{MAE}), and the maximum absolute error (\texttt{MaxAE}) of the fit are all zero. This is so because the number of clusters (122) is larger than the number of structures (50), thus, a perfect fit is possible. Conversely, the corresponding cross validation (CV) scores are different from zero. They express the ability of the built CE model to predict the properties of data not included in the fit \cite{Hastie2001}. The reason for a high CV score is either underfitting, or, what is the case now, overfitting. This can be avoided by searching on all possible subsets of clusters until finding the one for which the CV score is minimal. Such cluster-selection procedure, equivalent to regularizing with the $\ell_0$-norm (see Sec.~\ref{ssec:ceda}), is very demanding, since the number of  subsets increases exponentially, making the problem numerically tractable only for small clusters pools. Nonetheless, very good approximations to the optimal solution exist. 

Both the $\ell_0$ solution as well as approximations to it are available through the \texttt{ClustersSelector} class of \texttt{CELL}. Although this class can be used independently, CE models can be easily built through a helper class called \texttt{ModelBuilder}. It encapsulates both the cluster selection and the estimator construction. In this example, we use the  \texttt{ModelBuilder} class to find optimized models using two strategies. In the first one, the optimal solution is searched among sets of clusters of increasing radius and increasing number of points: For a given cluster to be present in a set, all other clusters with smaller radii and smaller numbers of points must be present as well. The other strategy makes use of the least absolute shrinkage operator (LASSO), which is a good approximation to the $\ell_0$ solution under certain conditions (see Sec.~\ref{ssec:ceda}). 

In the following listing, the code for an optimization with the first strategy is shown:

\noindent
\begin{minipage}{\linewidth}
\begin{lstlisting}[caption={Creation of a CE model with the helper class \texttt{ModelBuilder}.}, label={lst:modelbuilder}]
from clusterx.model import ModelBuilder
mb = ModelBuilder(
	selector_type="subsets_cv", 
	selector_opts={"clusters_sets":"size"}, 
	estimator_type="skl_Ridge", 
	estimator_opts={"alpha":1e-8, "fit_intercept":True})

ce_model = mb.build(sset, cpool, "e_ads")
ce_model.report_errors(sset)
\end{lstlisting}
\end{minipage}

\begin{figure}[htpb]
\centering
\includegraphics[]{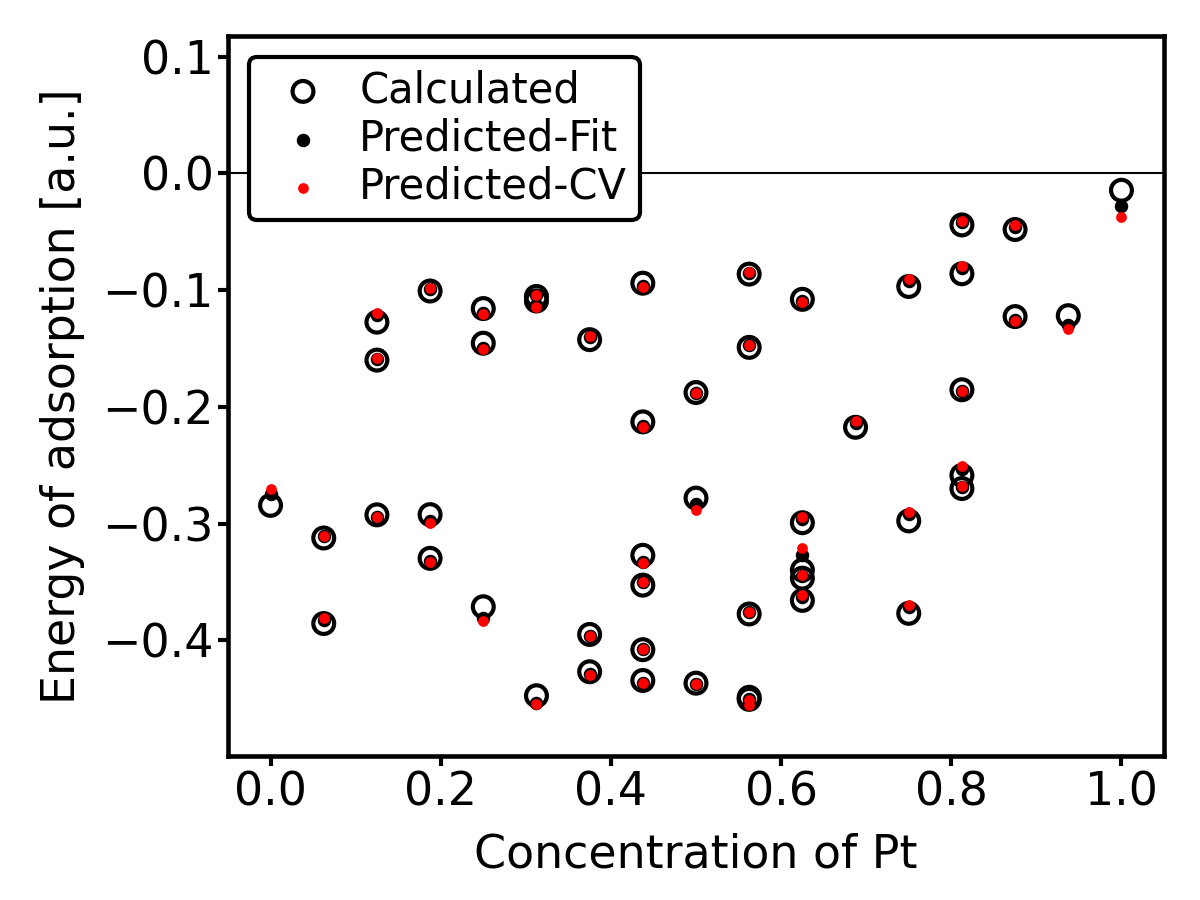}
\caption{Predicition of \textit{ab initio} adsorption energies and CVs, using the \texttt{ModelBuilder} class of \texttt{CELL} (Listing~\ref{lst:modelbuilder}).}
\label{fig:eads-size}
\end{figure}

\noindent
Here, the \texttt{ModelBuilder} class is imported (line 1) and instantiated in lines 2-6. Its initialization requires to specify (i) what strategy to employ to select the optimal model, and (ii) what estimator to use once the optimal set of clusters is determined. For (i), the arguments \texttt{selector\_type} and \texttt{selector\_opts} are set to \texttt{"subsets\_cv"}  and \texttt{{"clusters\_sets":"size"}}, respectively, indicating that the optimal solution will be searched among sets of clusters of increasing size, as explained above. For (ii), the values assigned to the arguments \texttt{estimator\_type} and \texttt{estimator\_opts} have the same meaning as in Listing~\ref{lst:ridge}. The keyword \texttt{"fit\_intercept"} is set to \texttt{True} (line 6), which amounts to add the empty cluster in the expansion (see Sec.~\ref{ssec:general}). Using this setup, the CE model is built in line 8, and the errors are reported in line 9. The output is shown in Fig.~\ref{fig:errsubsetcv}.
\begin{figure}[htpb]
\centering
\begin{BVerbatim}[fontsize=\small]
+-----------------------------------------------------------+
|                Report of Fit and CV scores                |
+-----------------------------------------------------------+
|                   |        Fit        |        CV         |
+-----------------------------------------------------------+
|       RMSE        |      0.00469      |      0.00677      |
|        MAE        |      0.00363      |      0.00505      |
|       MaxAE       |      0.01378      |      0.02306      |
+-----------------------------------------------------------+
\end{BVerbatim}
\caption{Output of Listing~\ref{lst:modelbuilder}.}
\label{fig:errsubsetcv}
\end{figure}
As compared to the previous CE model (Fig.~\ref{fig:outputridge}), here the fitting errors are not zero, however the generalization error (see column \texttt{CV}) is smaller, indicating that this model yields better predictions on new configurations, \ie configurations not contained in the training set \texttt{sset} used to build the model.  Figure~\ref{fig:eads-size} shows the predicted energies, both for the fit (black points) and CV (red points) as a function of Pt concentration. The figure is created as in Listing~\ref{lst:visua1}, by adding the argument \texttt{(..., cemodel = ce\_model, ...)} to the function call in line 2.

\begin{figure}[htpb]
\centering
\includegraphics[]{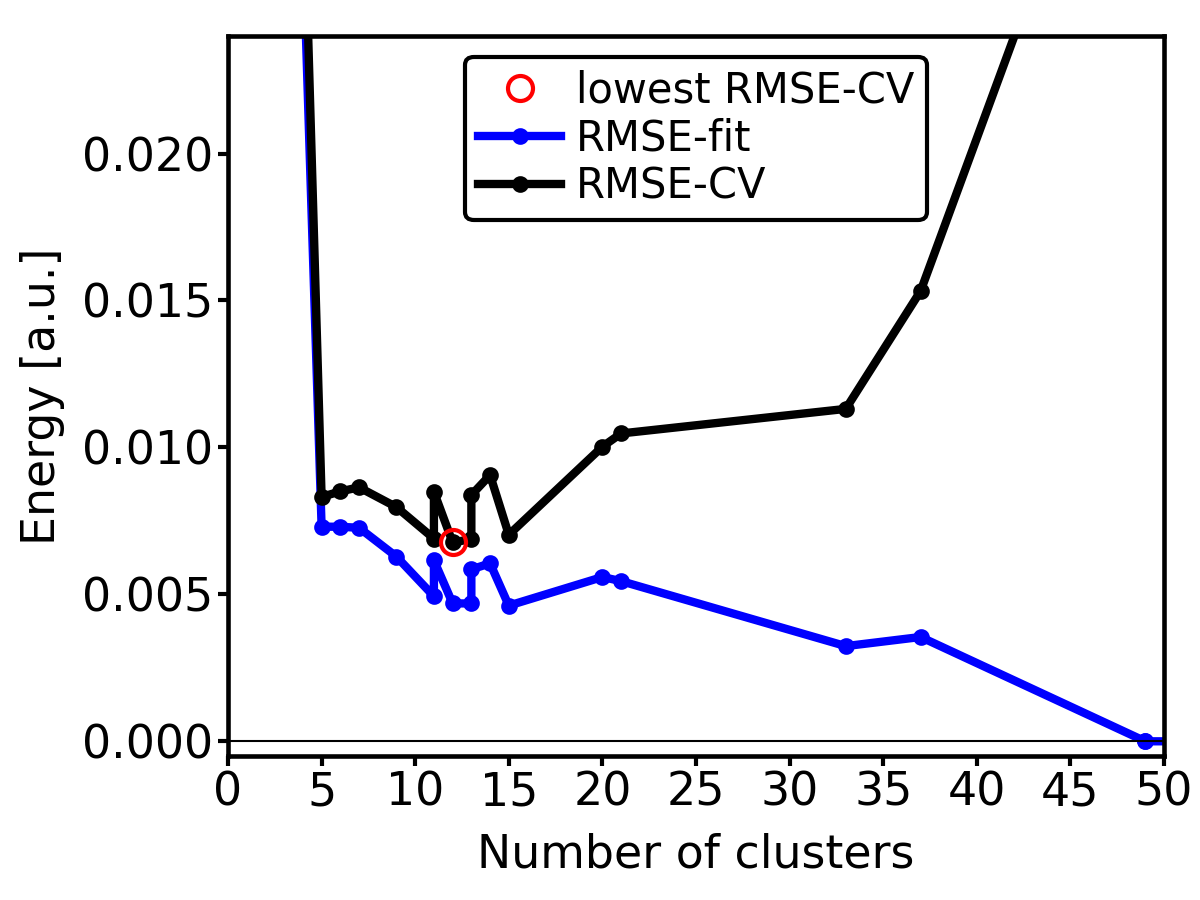} 
\caption{Optimization performed by the clusters selector.}
\label{fig:opt-size}
\end{figure}

It is interesting to consider in more detail the cluster optimization performed by the \texttt{ModelBuilder} class in Listing~\ref{lst:modelbuilder}. Figure~\ref{fig:opt-size} shows the RMSEs for both fit and CV for all sets of clusters considered, with the respective cardinality given in the abscissa. The optimal set, indicated with a red circle, contains 12 clusters.

The LASSO selector can be used by replacing lines 3 and 4 in Listing~\ref{lst:modelbuilder} with 

\begin{figure}[htpb]
\centering
\includegraphics[]{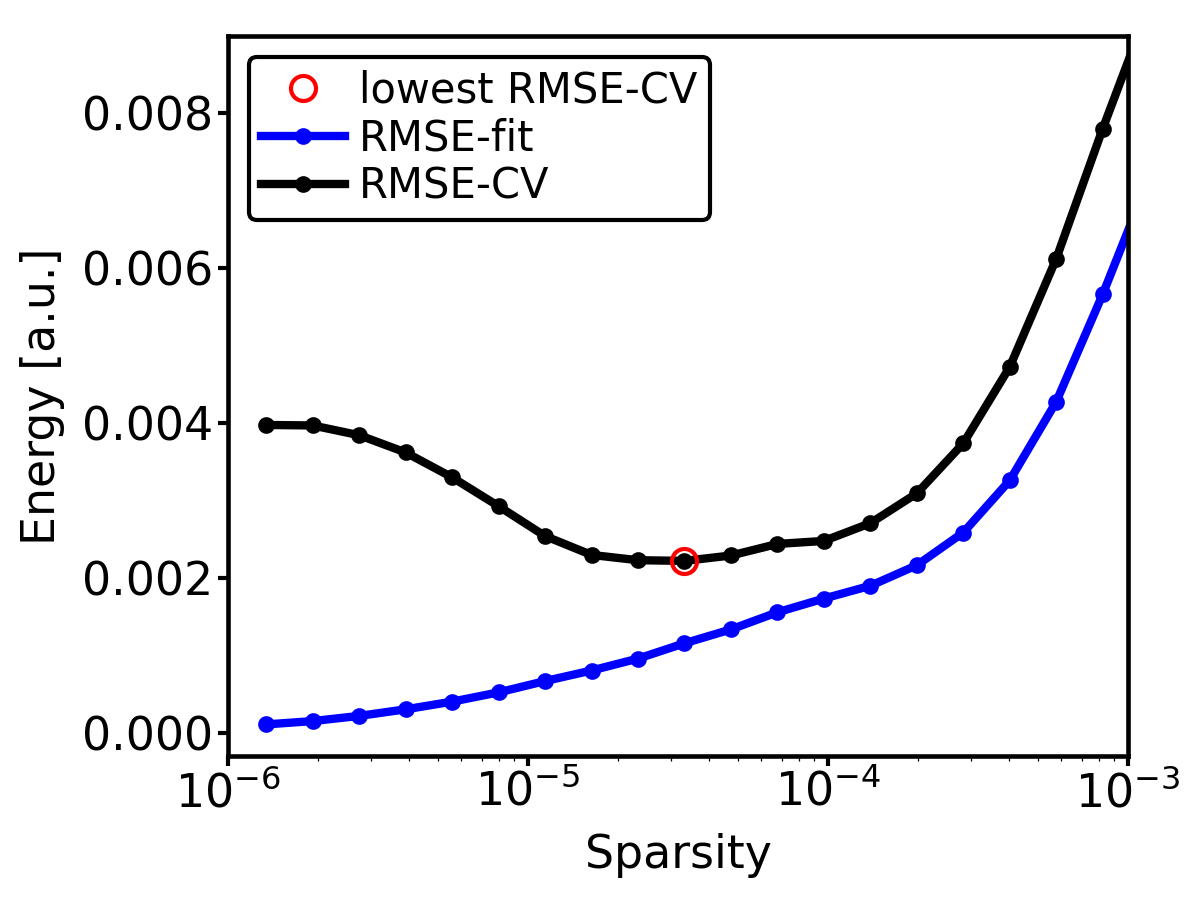}
\caption{Optimization performed by the \texttt{ClustersSelector} using LASSO.}
\label{fig:opt-lasso}
\end{figure}

\noindent
\begin{minipage}{\linewidth}
\begin{lstlisting}[caption={Creation of a CE model with LASSO using the \texttt{ModelBuilder} class.}, label={lst:lasso}]
selector_type="lasso_cv", 
selector_opts={'sparsity_max': 1e-2,'sparsity_min': 1e-6},
\end{lstlisting}
\end{minipage}
\noindent
In this case, the size of the cluster sets is controlled by the sparsity parameter $\lambda$ in the $\ell_1$ penalization term $\phi=\lambda \lVert \bm{{\cal J}}\rVert_1$ (see Sec.~\ref{ssec:ceda}): Larger values of $\lambda$ yield sparser models with smaller numbers of clusters and, conversely, smaller values of $\lambda$ yield larger cluster pools and eventually overfitted models. The interplay between sparsity and predictive power is shown in Fig.~\ref{fig:opt-lasso}. As expected, the RMSE of the fit decreases monotonously for decreasing sparsity, \ie the larger the cluster pool, the better the fit. 
Table \ref{tab:errors} shows the values of the RMSE errors for the three employed strategies for model construction.
\begin{table}[]
\centering
\begin{tabular}{|c|c|c|c|}
\hline
\multicolumn{1}{|l|}{} & 
\multicolumn{1}{c|}{Fit} & 
\multicolumn{1}{c|}{CV}& 
\multicolumn{1}{c|}{\#Clusters} \\ \hline
No optimization & 0.00000 & 0.01412 & 122\\ \hline
Subset-CV & 0.00469 & 0.00677 & 12\\ \hline
LASSO-CV & 0.00095 & 0.00183 & 22\\ \hline
\end{tabular}
\caption{Errors (RMSE) for models obtained with different strategies.}
\label{tab:errors}
\end{table}

\section{Thermodynamics\label{sec:thermo}}
The stability of multi-component alloys at finite temperatures is determined by the free energy, $F=U-TS$, where $U$ is the internal energy and $S$ the entropy. $S$ consists of at least three terms, namely, electronic, vibrational, and configurational contributions. The latter is usually approximated with the simple formula $S=-k_B\sum_i n_ix_i\log(x_i)$, with $n_i$ and $x_i$ being, respectively, the number of sites and fractional concentration of sublattice $i$, and $k_B$ the Boltzmann constant. Although being practical, it neglects interactions between substituent species, which can be crucial in determining the properties of complex alloys \cite{Troppenz2017}. In \texttt{CELL}, instead, the configurational entropy, including the effects of interactions, is accurately accounted for in the calculation of thermodynamic properties. This is achieved by employing different sampling methods, including the Metropolis Monte Carlo sampling and the Wang-Landau sampling, together with a CE model for the internal energy of the alloy. In this section, the application of these methods is demonstrated on the Pt/Cu(111) surface alloy using the CE model of Sec.~\ref{sec:ce}. 

As input, the sampling procedures require a CE model that predicts the energy of newly proposed structures during the sampling. The simulation cell and the thermodynamic ensemble must also be specified. In the canonical ensemble, for instance, it is necessary to provide the substituents' concentrations in the sublattices. A detailed documentation is provided in Ref.~\cite{Cell2019}. Listing~\ref{lst:sampini} 
\begin{figure}[htpb]
\begin{lstlisting}[caption={Initialization of a sampling procedure.}, label={lst:sampini}]
# Read the cluster expansion model for the absorption energy
from clusterx.model import Model
ce_model_ptcu = Model(filepath = "ce_model_ptcu.json")

# Create simulation cell 
from clusterx.super_cell import SuperCell
scell_ptcu = SuperCell(ce_model_ptcu.get_parent_lattice(),[[4,0],[-2,4]])

# Define concentration for the canonical sampling
nsubs={1:[4]} 
\end{lstlisting}
\end{figure}
shows the initialization of the \texttt{Model} object in line 3, corresponding to the adsorption energy, $E_{\rm ads}$, for the binary surface alloy (Eq.~(\ref{eq:eads}) with $n_O=0$). The simulation cell is created in line 7 by instantiating the \texttt{SuperCell} class that takes as arguments the parent lattice and a transformation matrix (\texttt{[[4,0],[-2,4]]}), which defines a rectangular simulation cell with 16 substitutional surface sites. In the examples demonstrated below, we perform samplings in the canonical ensemble, \ie with a fixed Pt concentration as specified in line 10. (Note that the dictionary key with value \texttt{1} refers to the sub-lattice corresponding to the top-most atomic layer of the Cu(111) surface.) Having 4 Pt atoms in 16 sites, leads to the stoichiometry of Cu$_3$Pt at the surface. 
  
With the CE model, the simulation cell, and the concentration already set up, we can start a canonical sampling. The Metropolis Monte-Carlo method is explained in Sec.~\ref{sec:mc}, and the Wang-Landau method in Sec.~\ref{sec:wl}.

\subsection{Metropolis Monte Carlo sampling\label{sec:mc}}

A widely used method to study finite-temperature properties in alloys is the Metropolis Monte-Carlo method \cite{Metropolis1953,Hastings1970}. In this sampling method, trial moves made by swapping two atoms from randomly chosen crystal sites, are accepted with the probability
\begin{equation}
P ( E_0 \rightarrow E_1) = \min \left[ \exp \left( - \frac{E_1-E_0}{k_{\rm B} T} \right) , 1 \right],
\end{equation} 
where $E_0$ is the energy of the initial structure, $E_1$  the energy after swapping the atoms, and $\exp ( - E/k_{\rm B} T )$ the Boltzmann probability distribution at temperature $T$. 

Listing~\ref{lst:mmc} demonstrates how to perform a Metropolis Monte-Carlo (MC) sampling with \texttt{CELL} and how to compute the specific heat at a given temperature after the sampling. First, the number of sampling steps \texttt{nmc} is specified in line 2. A set of temperatures, \texttt{temps}, is given in line 5. Here, the units of the temperature and $\mathtt{k_{\rm B}}$ have to be consistent with the energy units from the CE model, which is eV per substitutional site in our case. Thus, it is convenient to use eV/K for $\mathtt{k_{\rm B}}$, and K for temperature.

\begin{figure}[htpb]
\begin{lstlisting}[caption={Metropolis Monte-Carlo sampling with \texttt{CELL}.}, label={lst:mmc}]
# Define number of sampling steps 
nmc=100000
                             
# Define list of temperatures for the sampling (in Kelvin)
temps = range(1500,0,-50)

# Load Boltzmann constant in eV / K from ASE
from ase.units import kB

# Get number of Cu-Pt sites 
nsites = scell_ptcu.get_index()

# Perform Metropolis Monte Carlo samplings
from clusterx.thermodynamics.monte_carlo import MonteCarlo
mc = MonteCarlo(ce_model_ptcu, scell_ptcu, ensemble = "canonical", nsubs=nsubs)

cp_mc = []
for temp in temps:
	traj = mc.metropolis(nmc, [kB,temp,1/nsites], write_to_db = True)
	cp = traj.get_average_value("Cp", equilibration_steps = int(nmc/5))
	cp_mc.append(cp)

\end{lstlisting}
\end{figure}

\begin{figure}[htpb]
	\centering
	\includegraphics[scale=0.5]{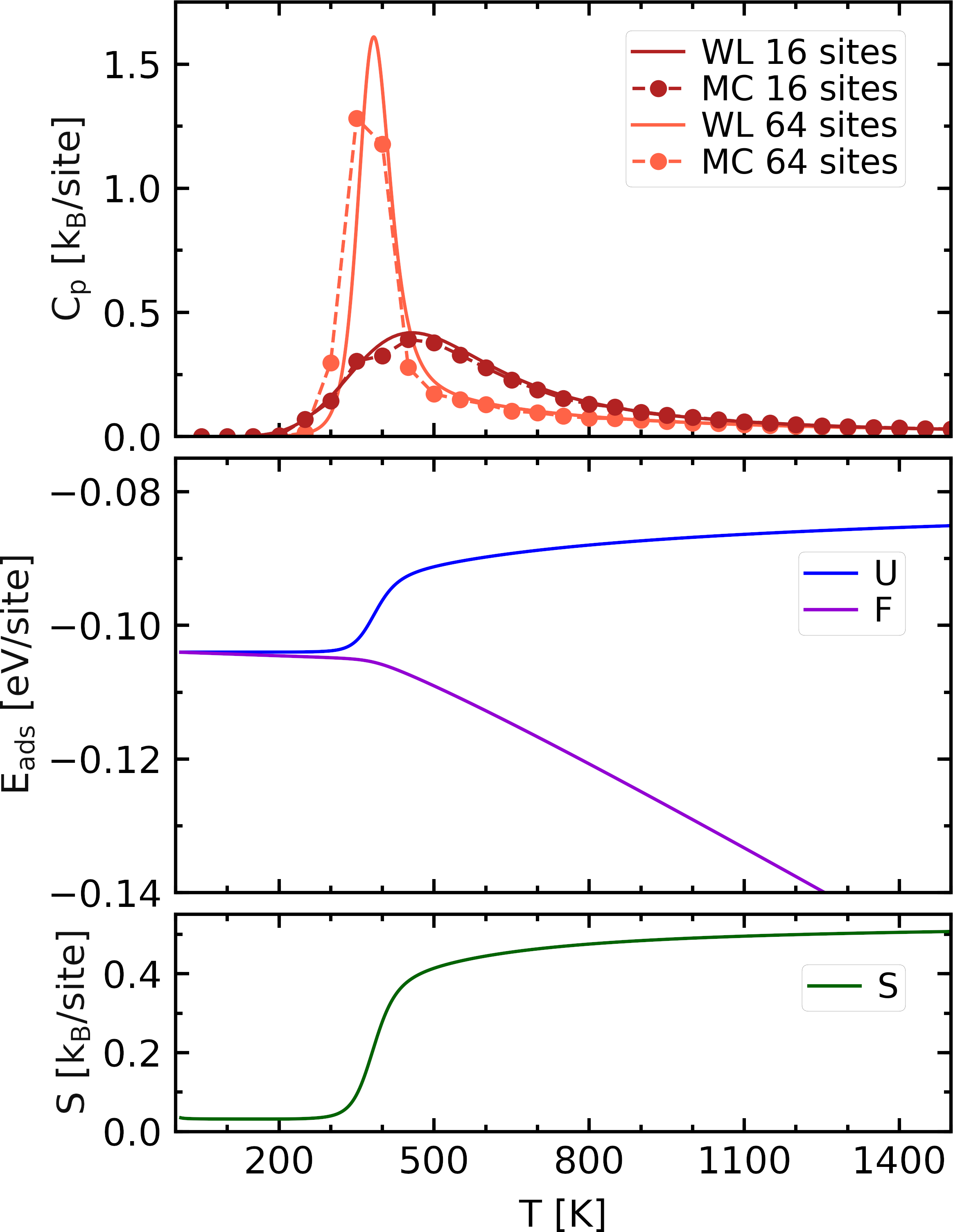}
	\caption{Thermodynamic properties of the Pt/Cu(111) surface alloy as a function of temperature. Top: Specific heat for simulation cells with 16 (dark red) and 64 surface sites (light red), respectively, obtained with MC sampling (dashed lines) and the WL method (solid lines). The middle panel shows the internal energy, $U$, and the free energy, $F$, for the simulation cell with 64 sites obtained by the WL method; the bottom panel the respective entropy, $S$.}
	\label{fig:thermo}
\end{figure}

The MC sampling requires a \texttt{MonteCarlo} object, which is created in line 15 by instantiating the \texttt{MonteCarlo} class of \texttt{CELL}. For the initialization, we indicate the CE model (\texttt{ce\_model\_ptcu}), the simulation cell (\texttt{scell\_ptcu}), the ensemble type (\texttt{"canonical"}), and the concentration (\texttt{nsubs}, see Listing~\ref{lst:sampini}). Using this instance (\texttt{mc}), an MC sampling is executed in line 19 by calling the method \texttt{metropolis} of \texttt{mc} for each temperature \texttt{temp} in the list \texttt{temps}. The number of sampling steps, \texttt{nmc}, is passed as argument. The product $\Pi$ of the elements in the second argument, \texttt{[kb,temp,1/nsites]}, enters the exponent of the acceptance probability as $\exp(-\Delta\hat{E}/\Pi)$, with $\Delta\hat{E}$ being the energy change predicted by the CE model. Since our CE model predicts energies per site, and the Boltzmann factor must be computed with total-energy changes, we include the factor \texttt{1/nsites} in the list. After successful completion of the sampling, a \texttt{MonteCarloTrajectory} object of \texttt{CELL} is returned and assigned to the variable \texttt{traj}. This object contains detailed information of the MC sampling trajectory, as \eg the energies of the visited structures and their configuration. Thus, any information of the MC trajectory can be read after the sampling procedure. For instance, by calling the method \texttt{get\_average\_value} of the \texttt{traj} object in line 20, we compute for each sampled temperature, the specific heat $C_p$, by passing the argument \texttt{"Cp"}. The result, shown in the top panel of Fig.~\ref{fig:thermo} (dark red dots connected by dashed lines) exhibits a maximum at $\sim 450$\,K. 

\begin{figure}[htpb]
	\centering
	\includegraphics[scale=0.35]{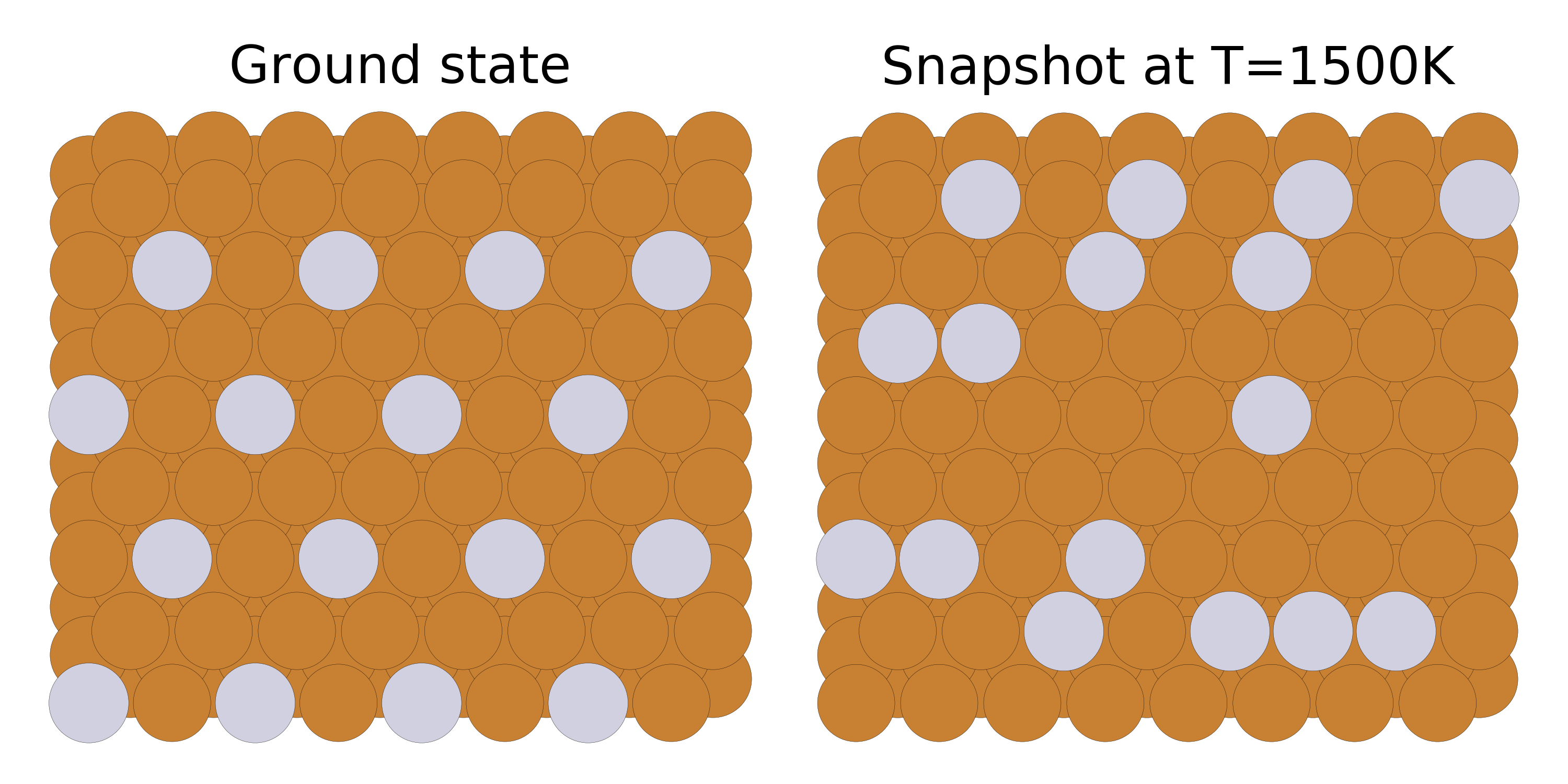}
	\caption{Ground-state structure (left) and snapshot at $T=1500\,$K (right) of the Pt/Cu(111) surface alloy in a simulation cell with 64 surface sites. Brown circles represent Cu atoms, gray circles Pt atoms.}
	\label{fig:thermo-structures}
\end{figure}

We can repeat the MC simulation with a larger simulation cell, \eg by passing the transformation matrix \texttt{[[8,0],[-4,8]]} in the initialization of the \texttt{SuperCell} object (line 7 of Listing~\ref{lst:sampini}), which produces a rectangular simulation cell of 64 sites. The resulting specific heat is shown in the top panel of Fig.~\ref{fig:thermo} (light red dots connected by dashed lines). As compared to the previous result with a smaller simulation cell, the peak becomes higher and narrower, and shifts to a lower temperature of around $350$\,K. This behavior signals an order-disorder phase transition. Indeed, from the MC trajectories, we see that the dominant configuration at low temperatures is the ground-state structure depicted in the left panel of Fig.~\ref{fig:thermo-structures}. It reveals a p(2x2) ordering of the Pt atoms, as expected from Ref.~\cite{Lucci2014}. For comparison, on the right side of Fig.~\ref{fig:thermo-structures}, a snapshot of the trajectory at 1500\,K is depicted, which looks rather disordered.

\subsection{Wang-Landau sampling\label{sec:wl}}
The Wang-Landau (WL) method \cite{Wang2001,Landau2004} aims at reducing the simulation effort by obtaining the configurational density of states, $g(E)$ --a temperature-independent quantity-- directly within the sampling procedure. Once $g(E)$ is known, the thermodynamic properties can be easily obtained at any temperature. This represents an enormous advantage as compared to MC simulations, particularly for quantities that require thermodynamic integration for its evaluation, like the free energy~\cite{Walle20022,Tuckerman2010}. In the WL approach, the latter is readily available, since the partition function $Z$ is directly obtained from $\int dE g(E)\exp(-E/k_BT)$. The WL algorithm is described in detail in Refs.~\cite{Wang2001,Landau2004}, and applications of this method can be found in Refs.~\cite{Borg2005,Khan2016,Troppenz2023}. In the following, its workflow is briefly explained, and its implementation in \texttt{CELL} is exemplified by the Pt/Cu(111) surface alloy. 

In the WL sampling, the energy space is sampled with the probability proportional to $1/g(E)$ which, for the exact $g(E)$, should produce a flat histogram $H(E)$ of the visited energies. At the start, however, $g(E)$ is unknown. Thus, the energy space is discretized into energy bins $E_i$, where each bin is assigned \textit{a priori} a uniform density of states $g(E_i)=1$. A newly proposed structure with energy $E_{\rm{new}}$ and density of states $g(E_{\rm{new}})$ (equal to 1 at the start) is accepted with the probability
\begin{equation}
P ( E_{\rm{old}} \rightarrow E_{\rm{new}}) = \min \left[ \frac{g(E_{\rm{old}})}{g(E_{\rm{new}})} , 1 \right] ,
\end{equation}
where $E_{\rm{old}}$ and $g(E_{\rm{old}})$ are the energy and density of states of the initial structure, respectively. If the trial structure is accepted (rejected), both $g(E)$ and $H(E)$ for the energy bin containing $E_{\rm{new}}$ ($E_{\rm{old}}$), are updated according to the following rule: $g(E)\rightarrow f\,g(E)$ and $H(E)\rightarrow H(E)+1$. With the multiplication factor $f$ kept fixed, the sampling is continued until the histogram satisfies a predetermined flatness condition. Then, the modification factor $f$ is reduced and the histogram restarted. The complete sampling procedure consists of a nested loop, with the inner loop generating the flat histogram and the outer loop reducing the modification factor. The accuracy of the final $g(E)$ is determined by the final flatness condition and modification factor. 

The WL algorithm is implemented within \texttt{CELL} in the \texttt{WangLandau} object and can be used as shown in Listing~\ref{lst:wl}. A \texttt{WangLandau} object is initialized in line 2 with the CE model of $E_{\rm{ads}}$, the simulation cell, and the given concentration, similar to the initialization of the \texttt{MonteCarlo} object in Listing~\ref{lst:mmc}. The WL sampling is then executed in line 4 by the method \texttt{wang\_landau\_sampling}. Here, \texttt{energy\_range} and  \texttt{energy\_bin\_width} define the energy window and bin width of the histogram, respectively. The modification factor $f$ and the flatness condition are predefined by default values that can be changed by the user. The initial modification factor $f=e$, is reduced by $f \rightarrow \sqrt{f}$ in each iteration of the outer loop, until the final modification factor is $f=\exp(10^{-3})$ (see argument \texttt{f\_range} in line 4). The condition for determining the flatness of the histogram is $\min[ H(E) ] > c \, \overline{H(E)}$, where the overline indicates the mean value, and $c$ is a real number. Starting with $c=0.5$ in the first iterations, it is then increased once every few iterations of the outer loop, until reaching $c=0.9$. More details and ways to control the default behavior can be found in Ref.~\cite{Cell2019}.

\begin{figure}[htpb]
\begin{lstlisting}[caption={Wang Landau sampling with \texttt{CELL.}}, label={lst:wl}]
from clusterx.thermodynamics.wang_landau import WangLandau
wl = WangLandau(ce_model_ptcu, scell_ptcu, nsubs)

cdos = wl.wang_landau_sampling(energy_range=[-0.103,0.0], energy_bin_width=0.002, f_range=[math.exp(1), math.exp(1e-3)])

# Temperature range for the calculation of the specific heat
temps = range(1500,0,-1)     
cp = cdos.calculate_thermodynamic_property(temps,"Cp")
\end{lstlisting}
\end{figure}

The algorithm returns a \texttt{ConfigurationalDensityOfStates} object, assigned to the variable \texttt{cdos} in line 4 of Listing~\ref{lst:wl}. Using this object, several thermodynamic quantities can be computed directly for an arbitrary number of temperatures. For instance, line 8 tells to compute the specific heat for the temperatures given in line 7. The result is shown in the top panel of Fig.~\ref{fig:thermo} (dark red solid line), together with that of the larger simulation cell (light red solid line) in comparison with the MC simulations. Since the evaluation of $C_p$ from $g(E)$ is computationally less demanding than the Metropolis Monte-Carlo method presented in the previous section, we can determine the transition temperature more accurately. In the lower panels of Fig.~\ref{fig:thermo}, the internal energy, $U$, the free energy, $F$, and the entropy, $S$, are presented. The entropy clearly increases with increasing temperature, indicative of an order-disorder phase transition. Note that the results presented here, can only show trends, but are not fully quantitative, since the CE model of the absorption energies is trained with model energies (see Sec.~\ref{sec:ce}.)

\section{Applications\label{sec:apps}}

In this section, we demonstrate the application of \texttt{CELL} to the binary alloy Si-Ge and to the clathrate Ba$_8$Al$_x$Si$_{46-x}$ as an example of a complex intermetallic alloy.

\subsection{Si-Ge alloy\label{ssec:sige}}
The binary alloy Si$_x$Ge$_{1-x}$ exhibits a miscibility gap, without forming ordered structures \cite{Qteish1988,Gironcoli1991}, \ie it has a strong tendency to separate into almost pure Si and Ge phases. This tendency decreases with increasing temperature until a critical temperature is reached beyond which the fully disordered phase is stable at any Si concentration. The theoretical description of this demixing transition has typically been done using Metropolis Monte Carlo simulations in the grand canonical ensemble~\cite{Gironcoli1991,Dunweg1993, Laradji1995, Walle2002, Garrity2019}. Here, we approach it using the Wang-Landau method in the canonical ensemble. This approach is interesting because it allows access to the phase-separation region, which is inaccessible in the grand canonical ensemble. We also describe the bowing of the lattice constant at different concentrations and temperatures. These results are compared with available experimental data.

Using a 16-atom supercell with lattice vectors $(a,a,0)$, $(a,0,a)$, and $(0,a,a)$, $a$ being the lattice constant of the parent diamond crystal, we generate a set of 43 structures with random configurations containing $n_{\text{Ge}}=0,1,2,...,16$ Ge substituents. The lattice constants are optimized, and the atomic positions relaxed until the forces are smaller than $5\times 10^{-3}$  eV/\AA. The \textit{ab initio} energies are calculated by density-functional theory (DFT) \cite{Hohenberg1964,Kohn1965} with the all-electron, full-potential electronic-structure code FHI-aims \cite{Blum2009,Knuth2015}. Exchange and correlation effects are treated within the generalized gradient approximation, employing PBEsol \cite{Perdew2008}. The basis set is determined by "tight" settings. A $10\times10\times10$ $\bm{k}$-point grid is used for integrations in the supercell Brillouin zone (BZ). 

\begin{figure}[htpb]
\centering
\includegraphics[]{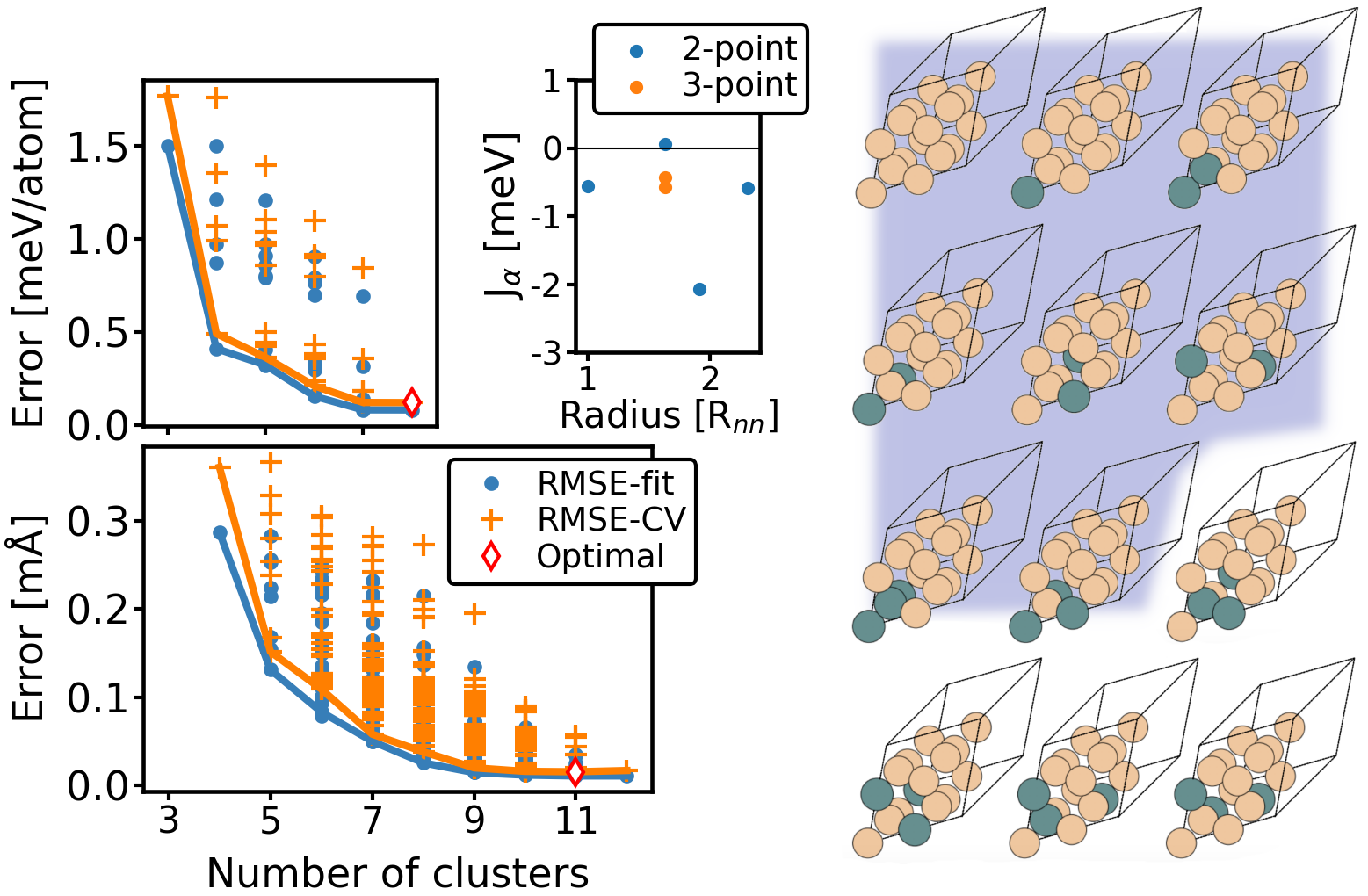}
\caption{CE models for SiGe. Upper left panel: Optimization of the CE model for the energy of mixing. Middle panel: Effective cluster interactions for the corresponding optimal CE model. Lower left panel: Optimization of the CE model for the lattice parameter. Right panel: Pool of clusters for building the CE models for the lattice constant and for the energy of mixing (blue background).}
\label{fig:sige-cemodels}
\end{figure}

The data generated in this way serve as the training set for CE models of the energy of mixing per atom, $E_{\text{mix}}$, and of the lattice parameter, $a_0$. The former is defined by $E_{\text{mix}}(x)=E(x)-\left[(1-x)E_{\text{Si}}+x E_{\text{Ge}} \right]$ with $E(x)$ being the total energy per atom of the compound with Ge concentration $x$, and $E_{\text{Si}}$ and $E_{\text{Ge}}$ the energies of the pristine Si and Ge solids, respectively. The CE models are built using a pool of clusters as shown in the right panel of Fig.~\ref{fig:sige-cemodels}, containing all clusters up to three points contained in the supercell: Besides the empty and a single one-point cluster, there are four 2-point clusters and six 3-point clusters. The cluster selection is performed by a combinatorial search on all possible subsets of clusters, with the condition that the first three (\ie the empty, the one-point, and the first-neighbor 2-point cluster) are included. The learning curves for the two CE models are shown in the left panels of Fig.~\ref{fig:sige-cemodels}. Here, the model which minimizes the RMSE-CV, marked by the red diamond, is selected. Every dot (plus sign) indicates the RMSE-fit (RMSE-CV) of a single trial model. The optimal model for $E_{\text{mix}}$ contains 8 clusters (blue background in the right panel), while for $a_0$ it contains 11 clusters. The solid orange line joins the models with optimal RMSE-CV as a function of the number of clusters, and the green solid line indicates the corresponding RMSE-fit for these models. While the latter decreases monotonously, the former may have a minimum as in the case of $a$ for 11 clusters, signaling overfitting for models with more clusters. The ECIs of the final CE model for $E_{\text{mix}}$ are displayed in the middle panel of Fig.~\ref{fig:sige-cemodels}. The cluster radii are given in units of the nearest-neighbor distance R$_\mathrm{nn}$, \ie the shortest distance between two Si atoms in the parent lattice. Most ECIs are negative, in accordance with the well known tendency of SiGe to phase-separate at low temperatures.

\begin{figure}[htpb]
\centering
\includegraphics[]{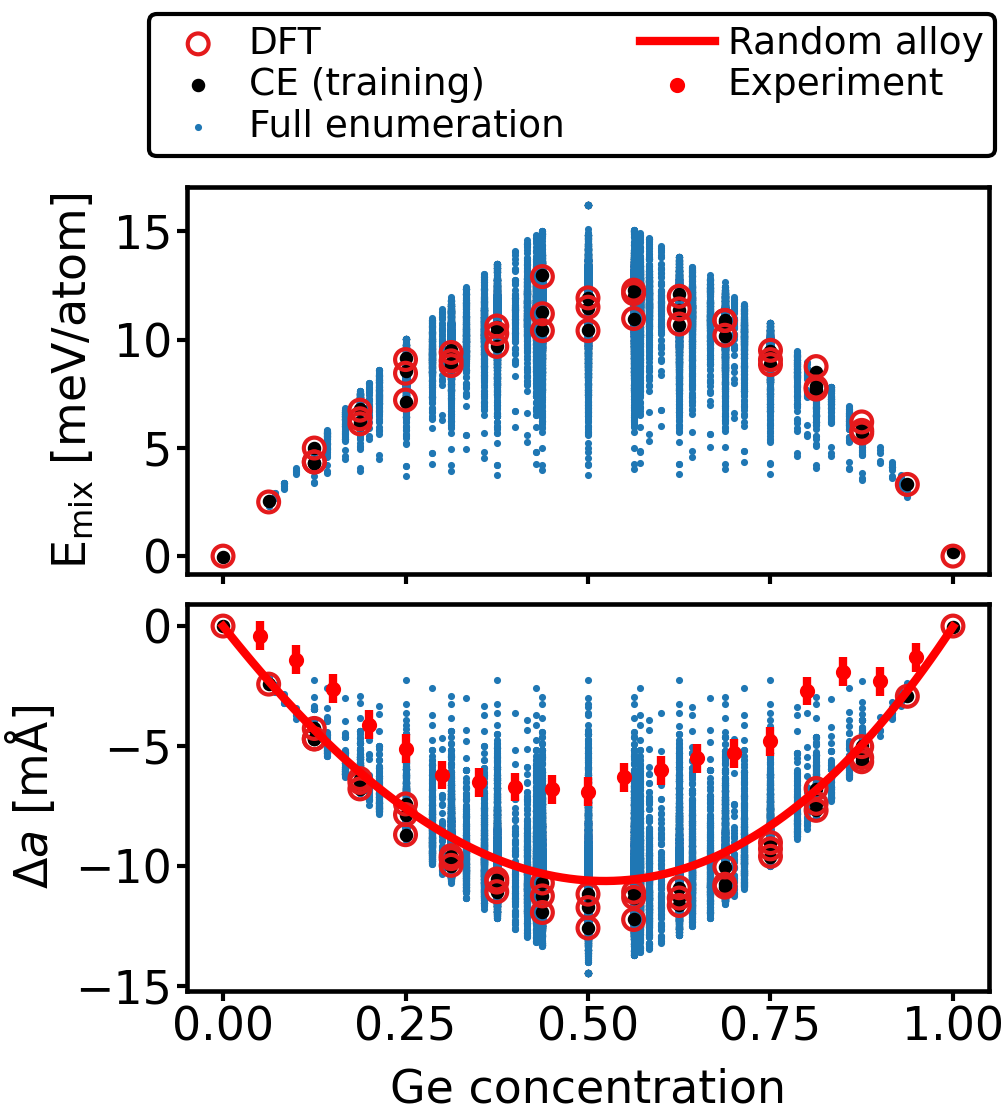}
\caption{DFT training data (red circles), CE predictions after training (black dots), and full structure enumerations (blue dots) for the CE of the energy of mixing (upper panel) and the lattice parameter (bottom panel) for the SiGe alloy as a function of Ge concentration. For the lattice constant, the predictions of the CE model for the perfectly random alloy are shown by a red solid line,  experimental values, taken from Ref.~\cite{Dismukes1964}, by red dots.}
\label{fig:sige_fullenum}
\end{figure}

The accuracy of these models becomes more clear in Fig.~\ref{fig:sige_fullenum} where excellent agreement between the DFT data (red circles) and the CE predictions (black dots) is observed. With these models, property predictions are made for all possible derivative structures of up to 16 atoms. The generation of the derivative structures is done with \texttt{CELL} using the algorithm of Ref.~\cite{Hart2008}. The resulting predictions are shown with blue dots. The fact that for all structures $E_{\text{mix}}\ge0$ means that at zero temperature, the system would favor a demixed state at all concentrations, without forming ordered structures. 

A negative bowing of the lattice constant is observed in the lower panel of Fig.~\ref{fig:sige_fullenum}, which indicates the departure from Vegard's law, which reads $\Delta a(x) = a(x)-\left[(1-x)a_{\text{Si}}+x \ \!a_{\text{Ge}}\right]$, with $a(x)$ being the lattice constant of a structure with Ge concentration $x$ and $a_{\text{Si}}$ and $a_{\text{Ge}}$ those of the pristine solids. For the perfectly random alloy, the cluster correlations can be computed analytically with \texttt{CELL}, so that the CE model can also be employed to compute the lattice parameter in this limiting case. This is shown by the solid red line, which yields $\Delta a$ values very close to the random structures used for training. The qualitative behavior is also similar to the experimental data, shown in the figure by red dots with error bars. Still, the experimental values are visibly smaller in magnitude than the prediction for the random alloy. This difference could in part be due to temperature-dependent configurational effects, as will be discussed below. 

\begin{figure}[htpb]
\centering
\includegraphics[]{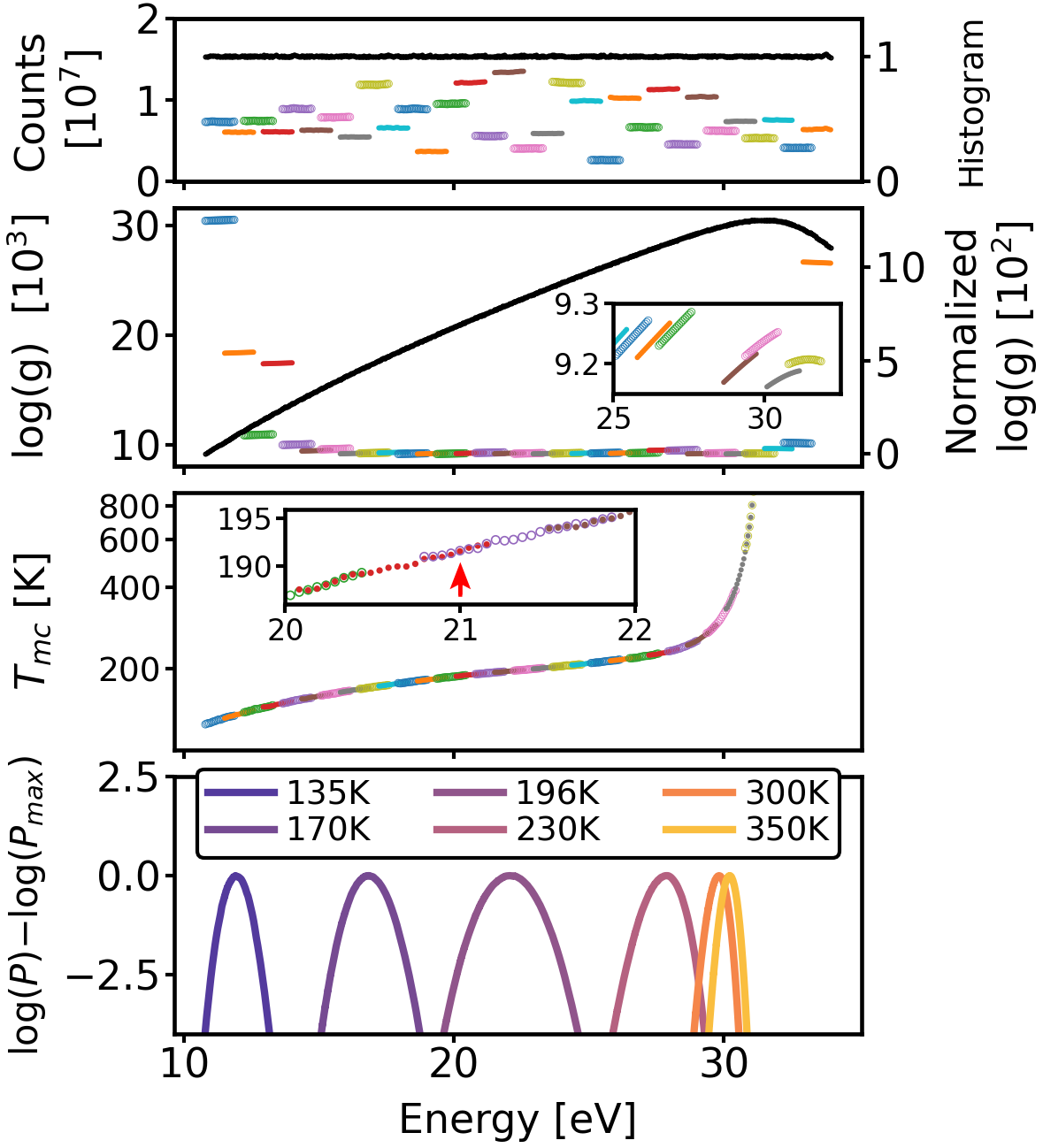}
\caption{Parallel Wang-Landau (WL) sampling with \texttt{CELL} for a supercell of 2744 atoms, using 32 processors. Top panel: WL histograms per CPU (colored dots and circles, left y-axis) and normalized histogram (black dots, right y-axis). Second panel: logarithm of the configurational density of states, log$(g)$, per CPU (colored dots and circles, left y-axis), and the normalized log$(g)$ (black dots, right y-axis). The inset shows a close up to highlight the energy dependency of the log$(g)$  in each process. Third panel: microcanonical temperature, $T_{mc}$, per processor. The arrow in the inset highlights the point of best overlap of $T_{mc}$'s. Bottom panel: logarithm of the canonical probability distribution, $\mathrm{log}(P)=\mathrm{log}(g)-\beta E$, for selected temperatures. All quantities are shown as a function of energy.}
\label{fig:sige_wl}
\end{figure}

Using these models, we study the temperature dependence of the demixing transition at 50\% Ge concentration for a cubic SiGe supercell containing 2744 atoms. Figure~\ref{fig:sige_wl} shows the results of a Wang-Landau sampling performed in parallel using 32 CPUs. Each process performs sampling in different overlapping energy ranges, as indicated by different colors in the first three panels of the figure. The first panel shows the converged WL histograms per process converged up to a flatness condition of $c=0.995$. Each histogram contains 22 bins with a bin width of 51 meV. In the last WL iteration, about $2\cdot10^8$ structures per energy range are visited. The normalized histograms, indicated by black dots, merge into a global flat histogram for the entire energy range considered. The latter is determined at the start of the simulation by considering the predicted energies of a fully demixed and a fully disordered structure, giving the minimum and maximum energies in the energy range, respectively. 

The logarithm of the configurational density of states, $log(g)$, is evaluated for all processes and converged up to a final modification factor $f$ satisfying $ln(f)=2^{-21}$. Upon reaching convergence, each process $i$ yields a configurational density of states $C_i \ \! g(E)$, with an arbitrary normalization $C_i$. This means that the $log(g)$ of contiguous processes $i$ and $i+1$ are shifted by an additive constant $log(C_i/C_{i+1})$. Therefore, to extract $g(E)$ with a common normalization factor, one has to shift the parts of $log(g)$ by appropriate additive constants. As described in Ref.~\cite{Vogel20141}, these are determined by finding the energies at which the microcanonical temperature $T_{mc}=(k_B\mathrm{d} ln(g)/\mathrm{d}E)^{-1}$ from adjacent energy ranges overlap best. The third panel of Fig.~\ref{fig:sige_wl} shows $T_{mc}$ for each process. Here, the red arrow in the inset illustrates the point of best overlap of $T_{mc}$'s for two contiguous ranges. The resulting $log(g(E))$ is shown by the black line in the second panel. It is normalized such that the  lowest-energy configuration has a degeneracy of one. It is to be noted though, that the results presented below do not depend on the chosen normalization.

From $log(g)$, one can easily evaluate the canonical probability distribution $P(E,T)$ at any temperature. This is useful for computing thermodynamic averages of different quantities, as will be shown below. In the fourth panel of Fig.~\ref{fig:sige_wl}, $log(P)-log(P_{max})$ is shown for various temperatures. As expected, the energy at which $P$ is maximum increases monotonically with temperature. At large temperatures, the maxima approach the limit of the fully random structure. This is why the maxima for $T=300K$ and $T=350K$, which are above the demixing transition temperature, are close to each other. Since, importantly, $log(P)$ always shows a single maximum, in agreement with Ref.~\cite{Dunweg1993}, the transition does not seem to be of first order, at least for the supercell sizes considered here. 

\begin{figure}[htpb]
\centering
\includegraphics[]{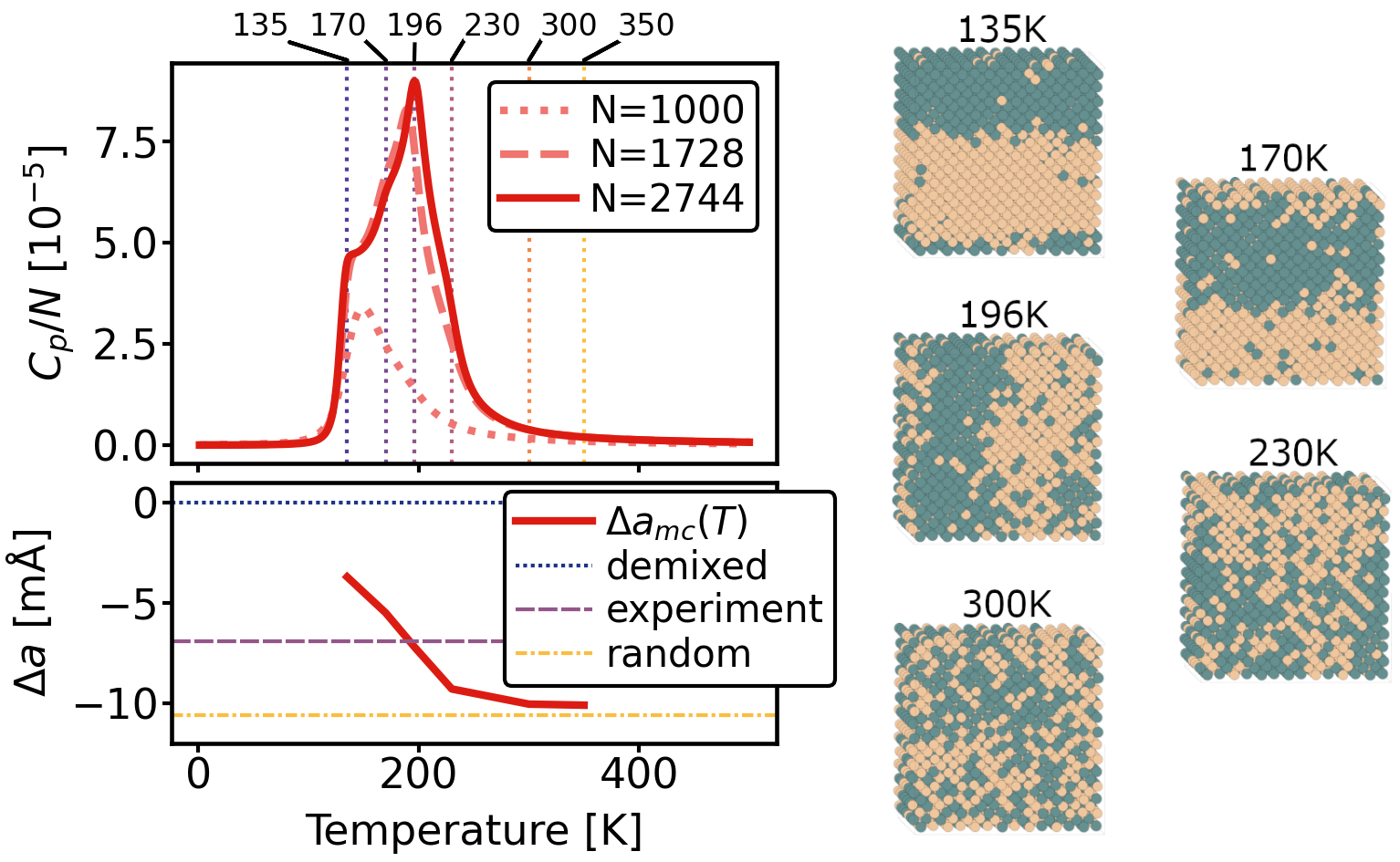}
\caption{Upper left panel: Specific heat at constant pressure, $C_p$, as a function of temperature for different numbers of atoms $N$. The thin dashed vertical lines and corresponding labels on top indicate temperatures at which $\Delta a$ is computed. Lower left panel: $\Delta a$ as a function of temperature; the horizontal lines mark the predictions for $\Delta a$ for the demixed state, the random state, and the experimental value from Ref.~\cite{Dismukes1964} at a Ge concentration of $51.3$ \%. Right panel: Samples from microcanonical sampling at energies that maximize $P(E,T)$ at the specified temperatures.}
\label{fig:sige_trans}
\end{figure}

The demixing transition is more closely analyzed in Fig.~\ref{fig:sige_trans}. Here, the specific heat at constant pressure, $C_p(T)=(\langle E^2 \rangle-\langle E \rangle^2)/k_BT^2$, with $\langle E^n \rangle=\int dE  \ \! E^n P(E,T)$, is computed using $P(E,T)$ from Fig.~\ref{fig:sige_wl}. The calculation is performed for three supercell sizes, containing 1000, 1728, and 2744 atoms, respectively. The specific heat peaks at a temperature that increases for increasing system size. For the largest system, the maximum is at about T=196K. At T$\sim$135K, $C_p$ displays a shoulder and at T$>$196K it decays monotonously to zero. It is interesting to explore what kind of structures are present at different temperatures, which are indicated by thin vertical dashed lines. These are obtained by taking a sample from a microcanonical sampling at a narrow energy range centered at the corresponding maxima of $P(E,T)$ (see bottom panel of Fig.~\ref{fig:sige_wl}) and are shown in the right panel of Fig.~\ref{fig:sige_trans}. At about 135K, the system is essentially separated into a pure Si and a pure Ge phase, with very little dissolution of one species into the bulk of the other. At 170K and 196K --the maximum of $C_p$-- this dissolution increases, indicating the proximity of the mixed state. Just above the maximum, at T=230K and in the region of small $C_p$ around T=300K, the structures appear fully mixed and increasingly disordered. These observations suggest a transition temperature of about $T_c\sim$200K. This value lies in between various values from previous simulations in the literature, such as 360K in Ref.~\cite{Qteish1988}, 170K in Refs.~\cite{Kelires1989,Gironcoli1991}, 320K in Ref.~\cite{Dunweg1993}, 247K in Ref.~\cite{Laradji1995}, or 325K in Ref.~\cite{Walle2002}.

 For each temperature shown in the upper left panel of Fig.~\ref{fig:sige_trans}, $\Delta a(T)$ is estimated in a \textit{single-shot} procedure. It consists of taking a single sample from a microcanonical sampling at an energy that maximizes $P(E,T)$ (as shown in the right panel of the figure) and then using the CE model for $a$ to predict $\Delta a$.  $\Delta a$ decreases monotonously with temperature as shown in the lower panel of the figure (solid red line). At around the lowest computed temperature (T=135K),  it approaches zero, which is the value predicted for the perfectly phase-separated SiGe alloy. Above $T_c$, at $T=300K$ and $350K$, $\Delta a$ is very close to the value predicted for the perfect disordered solid, as expected. At $T=196K$, our result is close to the experimental value reported in Ref.~\cite{Dismukes1964}. These findings imply a negative expansion coefficient in the transition region, purely due to configurational effects. This region is typically inaccessible to experiments, which usually deal with the homogeneous disordered phase above the transition temperature. In this case, the configuration being in the disordered state, does not change with temperature and due to anharmonicities of the crystal, the measured expansion coefficient is positive and estimated to be about 5.4 $\times 10^{-6} K^{-1}$ at a concentration of $51.3$ \% Ge~\cite{Dismukes1964}. Note that our calculations neglect anharmonic effects. These would presumably reduce the magnitude of the predicted $\Delta a$ for increasing temperatures.

\subsection{Si-based clathrate compounds\label{ssec:appclath}}
\begin{figure}[htpb] 
\centering
\includegraphics[width=0.55 \textwidth]{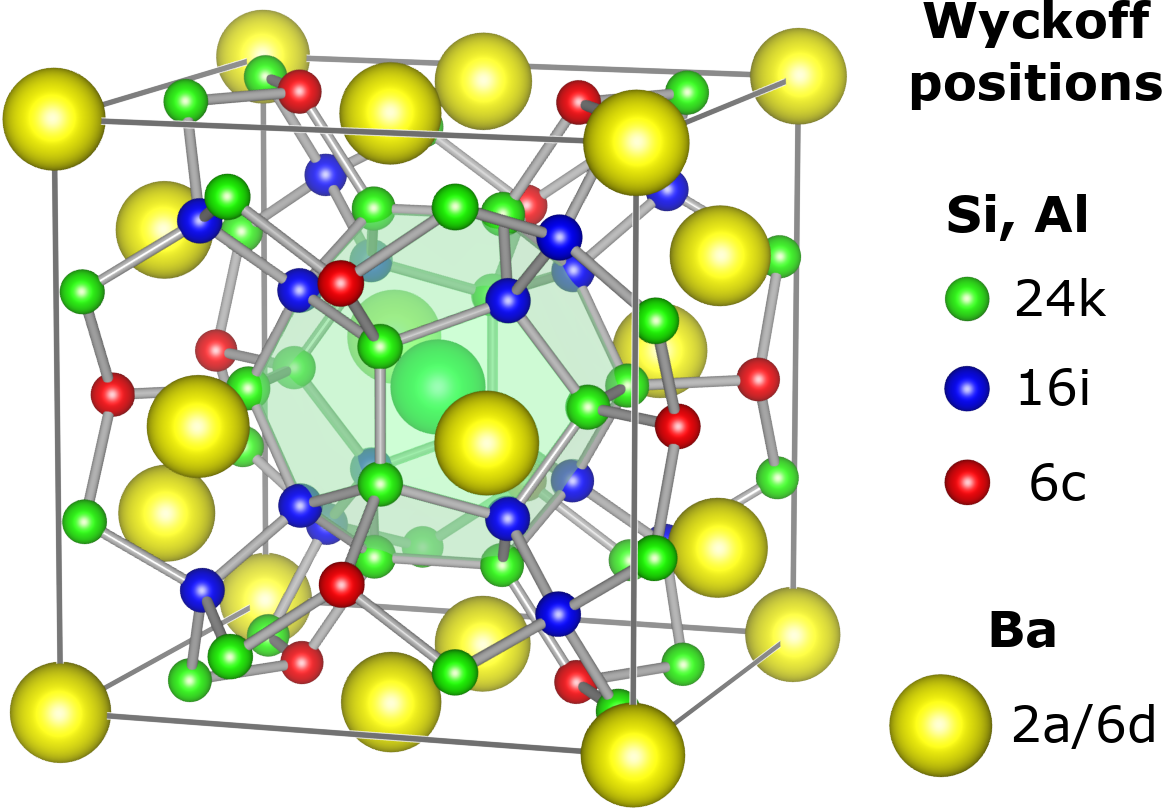}
\caption{Primitive cell of the type-I clathrate Ba$_8$Al$_x$Si$_{46-x}$, consisting of 54 atoms. The Si-Al host lattice (Wyckoff positions 24k, 16i, and 6c) forms a cage-like structure that encloses 8 Ba guest atoms (Wyckoff positions 2a and 6d). One of the cages is highlighted.
}
\label{fig:clath-struc}
\end{figure}

We now demonstrate the application of \texttt{CELL} to the study of a complex alloy, the intermetallic clathrate Ba$_8$Al$_x$Si$_{46-x}$. Intermetallic clathrates belong to the class of inclusion compounds in which the host lattice forms cages that can enclose guest species. Their low thermal conductivity together with their highly tunable electronic properties make them ideal candidates for thermoelectric applications \cite{Snyder2008}. The unit cell of Ba$_8$Al$_x$Si$_{46-x}$ is shown in Fig.~\ref{fig:clath-struc}. It contains 54 atoms, 46 of which are tetrahedrally bonded and form the host lattice of Si and Al atoms (Wyckoff sites 24k, 16i, and 6c). The 8 Ba atoms occupy the cages in Wyckoff sites 2a and 6d. 

\begin{figure}[htpb] 
\centering
\includegraphics[width=0.60 \textwidth]{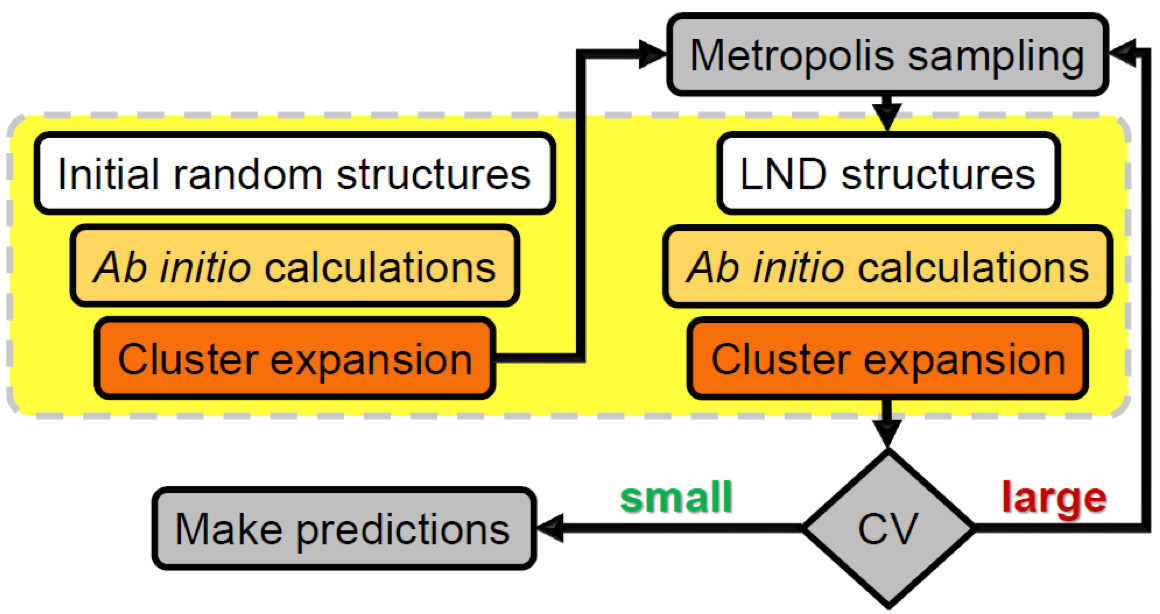}
\caption{Workflow assembled with \texttt{CELL}'s modular structure, for an iterative construction of a CE model, based on structural sampling instead of full enumeration.}
\label{fig:workflow}
\end{figure}

The material's properties, such as electronic structure and energy of formation, depend strongly on both the Al concentration $x$ and the crystal sites that they occupy, \ie the configuration \cite{Troppenz2017, Brorsson2021,Troppenz2023}. Therefore, it is of paramount importance to find the configuration of the ground-state structures (GSS), in order to understand the material's physical properties. This is, however, a formidable task. Due to the incredible number of possible configurations, \ie to arrange the Al atoms in the unit cell ($\sim$10$^{11}$ for $x=16$), approaches relying on a full enumeration of structures are infeasible. Therefore, in Ref.~\cite{Troppenz2017}, some of us have devised a special iterative approach to find the GSS and build an accurate CE model. The workflow, implemented in \texttt{CELL}, is depicted in Fig.~\ref{fig:workflow}. It is particularly useful for being applied to alloys with complex parent lattices, where a full structural enumeration is out of reach. First, the \textit{ab initio} energies of an initial set of random structures (left white box) are calculated (left yellow box), and an initial CE model is built with this data set (left orange box). Next, a configurational sampling is performed (upper gray box); the predicted structures with the lowest energies that are non-degenerate to those already present in the data set (labeled "LND structures" in the white box on the right), are added to the data set; and an improved CE model is determined (right orange box). The performance of the CE model is then evaluated through, \eg  cross validation. If the CV score is larger than the desired target precision, the CE model may be further improved by performing a new Metropolis sampling, as shown by the arrows. This is repeated until the CV score is smaller than the desired tolerance, and property predictions can be made. 

\begin{figure}[htpb]
\centering
\includegraphics[]{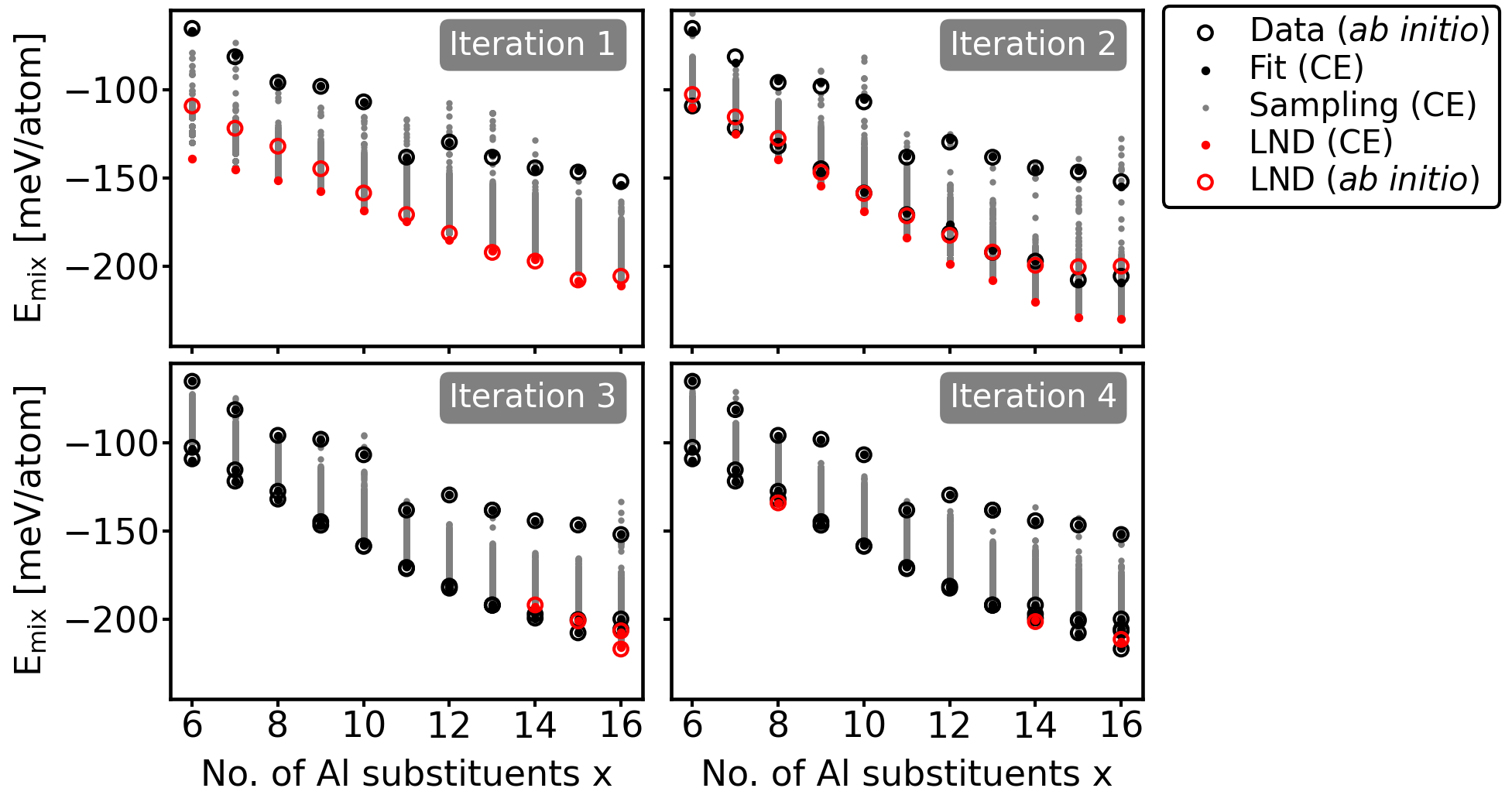}
\caption{Construction of a CE model and ground state search for the clathrate alloy Ba$_8$Al$_x$Si$_{46-x}$, using the workflow of Fig.~\ref{fig:workflow}. Black circles: \textit{ab initio} data used for training the CE model of the respective iteration. Black dots: CE predictions for the training data. Gray dots: Metropolis Monte Carlo samplings at 1000K with $5\times10^5$ steps per composition $x$. Red dots: CE predictions for the lowest non-degenerate (LND) structures found in each iteration. Red circles: \textit{ab initio} results for the LND structures.}
\label{fig:cla-iterations}
\end{figure}

Figure~\ref{fig:cla-iterations} shows the realization of this iterative cluster expansion for the clathrate alloy. The convergence of the CE model and the determination of the GSS are achieved in only 4 iterations. Here, the target accuracy of the CE model is 1meV/atom or less, since this allows one to correctly identify the GSS~\cite{Troppenz2017}. In the first iteration (top left), a set of 11 random structures with Al content in the range $x=6-16$ is created, and the energy of mixing of each structure is computed \textit{ab initio} with \texttt{exciting} using the functional PBEsol. \texttt{exciting} is a full-potential all-electron DFT package implementing the linearized augmented planewave + local-orbital method \cite{Gulans2014}. The energy of mixing per atom is defined by $E_{\text{mix}}(\sigma) = E(\sigma)-[\hat{E}_0(1-c)+\hat{E}_{46}c]$. Here, $E(\sigma)$ is the total energy per atom of the relaxed structure with atomic configuration $\sigma$, $\hat{E}_0$ ($\hat{E}_{46}$) is the predicted energy per atom of the hypothetical structure Ba$_8$Si$_{46}$ (Ba$_8$Al$_{46}$), and $c=x/46$. Computational details for the calculation of $E(\sigma)$ are given in Ref.~\cite{Troppenz2017}. The \textit{ab intio} computed energies are indicated by black circles in the figure. Following the workflow, an initial CE is constructed with these data. The predictions made with the corresponding CE model are represented by the black dots. Their RMSE-CV is 4.9 meV/atom. With this initial CE, a Metropolis sampling is performed for 1000K, with 5$\times10^5$ steps per composition  (gray dots) and the visited structures of lowest energy are identified (red dots). This step corresponds to the box "LND structures" of the workflow in Fig.~\ref{fig:workflow}. For these structures, \textit{ab initio} calculations are carried out (red circles). For iteration 1, the disagreement between the predictions for the LNDs and their \textit{ab initio} values for $x<12$ is big. In iteration 2 (upper right panel), the newly computed data from the previous iteration are added to the training set, which now consists of 22 data points shown as black circles. A new CE model is fitted to these data. Its RMSE-CV of 4.4meV/atom is slightly smaller than that of the first iteration but still well above the desired accuracy of 1meV/atom. Thus, according to the workflow, a new Metropolis sampling is performed (gray dots), LND structures are identified (red dots), and \textit{ab initio} calculations are performed for them (red circles). There is still a significant disagreement between predicted and \textit{ab initio} energies for the new LND structures, which increases with increasing $x$. In iteration 3 (lower left panel) again the data from the previous iteration are added, so that the training set (black circles) now consists of 33 structures. A new CE model trained with these data yields an RMSE-CV of 0.9 meV/atom, which is slightly below the desired accuracy threshold. Nonetheless, a new sampling is performed. The previously found GSS for $x<16$ are confirmed, and three new LNDs are added for $x>12$. For $x=16$, a new ground-state structure is found. These four data points (red circles) are added to the training set in the fourth and final iteration (lower right panel); the resulting model has an RMSE-CV of 0.8 meV/atom. A new sampling confirms the previously found GSS, except for two new ones found for $x=8$ and $x=14$. The latter are quasi-degenerate to the previously found ones. The 3 new data points from iteration 4 are added to the training set and a final CE model is fitted.

\begin{figure}[htpb]
\centering
\includegraphics[]{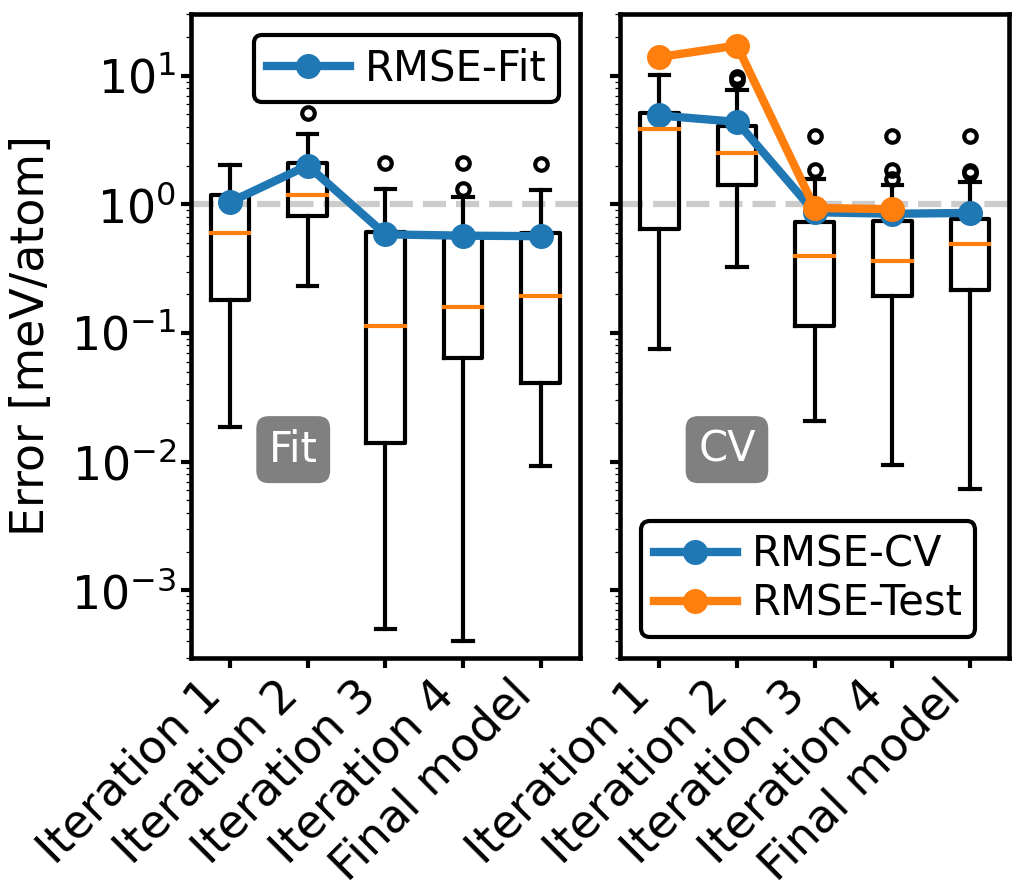}
\caption{CE model performance corresponding to the iterations presented in Fig.~\ref{fig:cla-iterations}. The box plots show the distribution of absolute errors, with the median indicated by thin orange lines and outliers by black circles. Left: fitting errors, right: generalization errors estimated with cross validation. RMSE-Test represents the errors from the red dots and circles of Fig.~\ref{fig:cla-iterations}. The gray dashed horizontal line indicates the target precision. }
\label{fig:cla-errors}
\end{figure}

A closer look at the model performance for the different iterations is provided in Fig.~\ref{fig:cla-errors}. Focusing first on the fitting errors (left panel), the RMSE is above the accuracy threshold in the first two iterations and then decreases to a value of around $0.6$meV/atom, below the desired accuracy. The median of the absolute errors follows a similar trend, but is considerably smaller in magnitude than the RMSE. The reason is that the latter is more sensitive to outliers. In the final model, a single outlier with an absolute error larger than 1meV/atom remains. For the generalization errors (right panel), a similar trend is observed. For the first two iterations, the RMSE-CV lies well above the accuracy threshold, but it stabilizes at a value of around $0.8$meV/atom for the remaining models. There is good agreement between the RMSE-CV and the RMSE-Test, indicating that the estimates of the generalization error given by the RMSE-CV are good. The RMSE-Test is obtained from the red dots and red circles in each iteration shown in Fig.~\ref{fig:cla-iterations}. Thus, it represents the error on totally unseen data, not even used for CV. 

The found CE model is able to make accurate predictions not only for GSS, but also for higher-energy structures. This allows a finite-temperature analysis using the thermodynamics modules of \texttt{CELL}. Such a study was performed in Ref.\cite{Troppenz2023}, where for the charge-balanced composition $x=16$, a temperature-driven semiconductor-to-metal transition was found. The latter is accompanied by a partial order-disorder structural transition.

\section{Conclusions\label{sec:con}}
We have described in detail the Python package \texttt{CELL} for cluster expansion and for statistical thermodynamics. \texttt{CELL} provides a modular framework that allows for customized CE model building and integration into workflows. This paves the way to address a wide range of problems, as illustrated by various applications. The first one, has demonstrated the ability of \texttt{CELL} to create the CE model of a complex surface system, O-Pt/Cu(111). This system is characterized by the interplay of two binary sublattices, one describing Pt/Cu surface alloying and the other the oxygen surface adsorption. Such systems are of great interest, for instance, for applications in catalysis. The characterization of the structure at finite temperature revealed a temperature-driven order-disorder transition. The found ordered low-temperature phase is in agreement with experimental findings.

The second example showed the application of \texttt{CELL} to the binary semiconducting alloy Si-Ge. Using a combinatorial approach to model selection, accurate CE models for the mixing energy and lattice parameters were generated. The prediction of the energy of mixing for all derivative structures of up to 16 atoms confirmed the tendency of Si-Ge to separate into pure Si and Ge phases. Consideration of the fully random alloy together with the CE model for the lattice constant, revealed a negative bowing of the lattice parameter, in agreement with experiments. We have characterized the demixing transition of Si-Ge in terms of the configurational density of states, the microcanonical temperature, the canonical probability distribution, the specific heat, and the thermal expansion. Our analysis, performed in the canonical ensemble, both contrasts with and nicely complements previous studies in the literature using the grand canonical ensemble. The latter yields homogeneous phases but prevents access to the phase-separated state, which is accessible in our study. In this example, we have showcased the parallel execution of \texttt{CELL}'s with supercells containing thousands of atoms. Inspection of structures from microcanonical sampling yielded a demixing transition temperature of around 200K, which is in the range of values reported in the literature. 

The third and last example concerned the construction of a CE model for the energy of the complex clathrate alloy  Ba$_8$Al$_x$Si$_{46-x}$. This material, with its 54 atoms in the unit cell, posed a formidable challenge to the tasks of building accurate CE models and finding the lowest-energy structures. We have addressed this problem by creating an iterative workflow that efficiently solves both problems simultaneously. The workflow uses continuously improved CE models to perform configurational sampling. These are used to identify low-energy structures that are iteratively added to the training set. With only four iterations, requiring the \textit{ab initio} calculation of just 40 structures, all ground states in the range $x=6-16$ are found. Analysis of the model's performance in terms of fitting and generalization errors revealed that convergence is already achieved by the third iteration. 

In summary, \texttt{CELL} provides a comprehensive approach to cluster expansion, covering all aspects of model construction and thermodynamical analysis. \texttt{CELL} offers its users an efficient way to build CE models using machine-learning techniques and leveraging parallelization, seamlessly integrating with Python code and facilitating the interaction with \textit{ab initio} packages. 

\section{Data and software availability\label{sec:avail}}
\texttt{CELL} is available at the Python Package Index (PyPI) repository \url{https://pypi.org/project/clusterX/}. Documentation can be found at \url{https://sol.physik.hu-berlin.de/cell}. It includes installation instructions, the API, and tutorials. Jupyter notebooks and Python scripts for reproducing the CE of O-Pt/Cu(111) of Sec.~\ref{sec:ce} are available on Github (\url{git@github.com:srigamonti/optcu.git}). The \textit{ab initio} data for Si-Ge (Sec.~\ref{ssec:sige}) and Ba$_8$Al$_x$Si$_{46-x}$ (Sec.~\ref{ssec:appclath})  are available in NOMAD \cite{Draxl2019}, DOI \url{https://dx.doi.org/10.17172/NOMAD/2023.10.24-3} and DOI \url{https://dx.doi.org/10.17172/NOMAD/2023.10.24-1}, respectively.

\section*{Acknowledgements}
We thank Luca Ghrinighelli for fruitful discussions on model selection and sampling methods and Matthias Scheffler for drawing our attention to the Wang-Landau method. This work received partial funding from the German Research Foundation (DFG) through the CRC 1404 (FONDA), project 414984028, and the NFDI consortium FAIRmat, project 460197019; the Max Planck Research Network BiGmax, and the European Union's Horizon 2020 research and innovation program under the grant agreement N$^{\circ}$ 951786 (NOMAD CoE). \\

\bibliographystyle{apsrev4-1}
\bibliography{cell-paper}

\end{document}